\documentclass[journal=jacsat,manuscript=article]{achemso}

\usepackage[version=3]{mhchem} 
\usepackage[font=footnotesize,labelfont=bf]{caption}
\usepackage{color}
\usepackage{soul}
\usepackage{lineno}
\usepackage[utf8]{inputenc}

\author{Filippo Franceschini}
\email{filippo.franceschini@kuleuven.be}
\affiliation[KULF]{Department of Physics and Astronomy, KU Leuven, Leuven}
\author{Catarina Fernandes}
\affiliation{Department of Electrical Engineering, KU Leuven, Leuven}
\author{Koen Schouteden}
\affiliation[KULF]{Department of Physics and Astronomy, KU Leuven, Leuven}
\author{Jon Ustarroz}
\affiliation{Chemistry of Surfaces, Interfaces and Nanomaterials (ChemSIN), Université libre de Bruxelles, Brussels}
\alsoaffiliation{Electrochemical and Surface Engineering (SURF) Department, Université libre de Bruxelles, Brussels}
\author{Jean Pierre Locquet}
\author{Irene Taurino}
\affiliation[KULF]{Department of Physics and Astronomy, KU Leuven, Leuven}
\alsoaffiliation{Department of Electrical Engineering, KU Leuven, Leuven}
\title[title]
  {Tailoring the Glucose Oxidation Activity of Anodized Copper Films on Microfabricated Platforms}


\begin{document}


\begin{abstract}
Glucose oxidation is a fundamental reaction in biosensing and energy conversion, and anodized copper electrodes have emerged as promising catalysts for its enhancement. This paper systematically investigates the link between anodization parameters and glucose oxidation on thin copper films, unraveling crucial insights into their optimization. By careful control of the anodization parameters, distinct species can be favoured, such as $\mathrm{Cu_2O}$, $\mathrm{Cu(OH)_2}$, and CuO, with varying catalytic activities.  Applying a polarization in 1 M KOH at 0 V (vs Ag\textbar AgCl) results in the formation of a highly active surface CuO layer, which delivers significant performance improvements compared to the bare copper electrode. Namely, a 55 \% sensitivity gain in the 0.1 mM - 0.5 mM range, and a remarkable 73 \% gain in the 0.75 mM - 2mM range. Furthermore, the manufactured electrodes display an extremely low limit of detection — only 0.004 mM. Such an exceptional result can be ascribed to the scalable and reproducible manufacturing done directly on cleanroom-compatible platforms. These insights not only clarify the effect of thin copper film anodization on glucose oxidation, but also chart a practical path towards improving the real-world efficiency of copper-based integrated systems for biosensing and energy conversion.
\end{abstract}

\section{Introduction}
Glucose oxidation is crucial in various domains such as biosensing, biofuel cells, and energy storage. Among these, diabetes monitoring currently has the greatest economic and societal relevance, with a market value of \$11.71 billion in 2021\cite{Das_Nayak_Krishnaswamy_Kumar_Bhat_2022} and more than 537 M people affected by the disease\cite{WHO}. Enzymes, such as glucose oxidase, are commonly used in glucose sensing technologies, exhibiting good sensitivity and selectivity\cite{Teymourian_Barfidokht_Wang_2020}. However, their storage and usage conditions require careful control due to their sensitivity to pH and temperature\cite{Wang_He_Wang_Gu_Huang_Fang_Geng_Zhang_2013}. The limited stability of these sensors, typically lasting a few weeks, is considered their primary weakness\cite{Wei_Qiao_Zhao_Liang_Li_Luo_Lu_Shi_Lu_Sun_2020,Jing_Chang_Chen_Liu_2022}. Compared to their enzymatic counterparts, non enzymatic sensors not only are more durable, but they also exhibit greater sensitivities and can be more easily miniaturized, as they are not limited by macro-sized molecules\cite{Aun_Salleh_Ali_Manan_2021}. Over the last decades, a plethora of non enzymatic glucose electrocatalysts has been developed and investigated. Early works highlighted how noble metals (e.g., Pt\cite{Taurino_Sanzò_Antiochia_Tortolini_Mazzei_Favero_Micheli_Carrara_2016,Vassilyev_Khazova_Nikolaeva_1985}, Pd\cite{Yang_Zhang_Lan_Chen_Chen_Zeng_Jiang_2014}, and Au\cite{Pasta_Ruffo_Falletta_Mari_Pina_2010,Hsiao_Adžić_Yeager_1996}) can directly oxidize glucose in neutral solutions. Nevertheless, the high cost and low resistance towards chloride poisoning is still limiting their application for glucose sensing\cite{Bagotzky_Vassilyev_Weber_Pirtskhalava_1970,Pasta_Mantia_Cui_2010}. As a result, transition metal oxides (e.g., $\mathrm{Ni(OH)_2}$\cite{Franceschini_Taurino_2022}, $\mathrm{Cu_xO}$\cite{Naikoo2022}, $\mathrm{Co_3O_4}$\cite{Ding_Wang_Su_Bellagamba_Zhang_Lei_2010}) have also been explored, sometimes with carbon-based materials as supports (e.g., carbon nanotubes\cite{Gupta_Gupta_Huseinov_Rahm_Gazica_Alvarez_2021} and boron-doped diamond\cite{Gao_Zhang_Zhao_Yan_Han_Li_Liu_Li_Liu_2022}). Among these, copper oxides are particularly appealing due to their exceptional catalytic activity and resistance to poisoning\cite{Li_Su_Zhang_Lv_Xia_Wang_2010,Colon_Dadoo_Zare_1993,Xie_Huber_1991}. Substantial research effort has been devoted to synthesizing $\mathrm{Cu_xO}$-based electrodes in diverse forms, such as nanoparticles\cite{Singh_Hazarika_Dutta_Bhuyan_Bhuyan_2022}, nanocubes\cite{Jiménez-Rodríguez_Sotelo_Martínez_Huttel_González_Mayoral_García-Martín_Videa_Cholula-Díaz_2021}, nanoneedles\cite{Yang_Li_Wang_Qiao_Qu_2017}, nanowires\cite{Zhang_Su_Manuzzi_Monteros_Jia_Huo_Hou_Lei_2012}, and nanoﬂowers\cite{nanoflowers}. A common fabrication approach involves directly immobilizing copper oxide catalysts onto a supporting material such as glassy carbon. However, the weak adhesion of $\mathrm{Cu_xO}$ structures to the surface often leads to unreliable results. To address this issue, the use of a polymer binder, such as Nafion\textsuperscript{\tiny\textregistered}, has been proposed\cite{Meher_Rao_2013,Ashok_Kumar_Tarlochan_2019}. Alternatively, copper oxides can be synthesized on the surface of Cu substrates through wet-chemical methods \cite{Zhang_Su_Manuzzi_Monteros_Jia_Huo_Hou_Lei_2012} or thermal oxidation \cite{Cao_2022}. Although these techniques offer improved adhesion of the oxide to the substrate surface, achieving large-scale and reproducible synthesis of uniform materials remains challenging. In contrast, electrochemical oxidation methods, such as potentiostatic, potentiodynamic or galvanostatic anodization, have been employed as direct fabrication methods for highly sensitive and stable $\mathrm{Cu_xO}$-based glucose sensors\cite{Babu_Ramachandran_2010,Dhara_Stanley_Ramachandran_Nair_Babu_2016,wang2012facile,Fan_Weng_Lee_Liao_2019,Stepniowski2020}. Compared to traditional synthesis methods, anodization offers several advantages. It is a relatively simple and low-cost technique that can be conducted under ambient conditions without the need for high temperatures or toxic chemicals. Moreover, it allows for precise control over the thickness and morphology of the surface oxide/hydroxide layer, enabling the reproducible design of remarkably active surfaces. Depending on the chosen electrolyte and applied potential/current density, different morphologies can be produced. Notable mentions include the galvanostatic production of nanowires by Wu et al. \cite{Wu_Bai_Zhang_Chen_Shi_2005} in KOH solution, the controlled nanowire diameter by potentiostatic anodization (10-50 V) in carbonate solution by St\c{e}pniowski \cite{stepniowski}, nanoporous structures in acqueous/nonacqueous electrolytes at high voltage ($\approx 10V$) by Allam et al. \cite{Allam_Grimes_2011}. Although promising, the research has solely investigated the properties of thick copper foils which don't necessarily translate to microfabricated sensors. This discrepancy arises from the generally higher resistivity of thin films compared to bulk electrodes \cite{Cui_Hutt_Conway_2012,Zhang_Brongersma_Richard_Brijs_Palmans_Froyen_Maex_2004}; a challenge further exacerbated by the complexity of removing $\mathrm{O_2}$ impurities in widespread PVD sytems such as sputtering\cite{Schmiedl_Wissmann_Finzel_2008,Su_2016,Chandra}. Moreover, the lack of consensus on the mechanism glucose oxidation on copper combined with the mixed phase composition of most anodized surfaces, requires the validation of each technique for glucose oxidation. To address this issue, researchers have directly evaluated the ability of differently anodized copper foils to oxidize glucose. Nonetheless, most of these studies \cite{Babu_Ramachandran_2010,Fan_Weng_Lee_Liao_2019,Li_Wei_Fu_Wang_Chen_Li_Li_Song_2014,Xu_Yang_Liu_Liu_Sun_2013} have focused on specific sets of anodization parameters without providing a justification for their selection. Recently, the group of Noda\cite{Anantharaj_Sugime_Yamaoka_Noda_2021} investigated the effect of the hydroxide ion strength in the anodization solution for methanol oxidation. However intriguing, their work studied the effect of the $\mathrm{OH^-}$ concentration at a single potential, thus leaving many questions still unanswered. This lack of exhaustive understanding and rationalization hinders the progress of $\mathrm{Cu_xO}$-based catalysts. Therefore, the objective of this paper is to provide a comprehensive outline of the potentiostatic anodization process applied to thin film copper electrodes, specifically for glucose oxidation applications.
By increasing in small steps the anodization potential together with the KOH concentration, we aim to evaluate the glucose sensitivity for each individual condition. In this way, we can establish a distinct correlation encompassing the cyclic voltammetric curve, overall sensitivity, and the surface physico-chemical properties generated through each set of anodization parameters.
The results of this work will help to fill the knowledge gaps and facilitate the development of highly efficient and integrated $\mathrm{Cu_xO}$-based glucose sensors and electrocatalysts more broadly.
\section{Experimental section}
\subsection{Electrode preparation}
In this work, 7x7 mm Cu/Pt/glass electrodes were fabricated starting from 4” borosilicate glass wafers. Through RF magnetron sputtering, a 300nm thick layer of Pt was deposited, acting as a current collector. Pt was chosen over Au due to its greater compatibility with the fabrication of integrated electrochemical sensors, where the counter electrode is commonly Pt. Afterwards, 300 nm layer of Cu was deposited on top of the Pt and through careful control of base pressure ($2.5*10^{-5}$ mBar) and working pressure ($1*10^{-2}$ mBar) a well adherent and non-oxidized layer was obtained, as confirmed by GI-XRD (\textbf{S \ref{fig:XRD}}). The deposition parameters are reported in \textbf{S Table \ref{tab:fab}}. The thickness of the sputtered Cu thin films was assessed using a Dektak Profilometer (Veeco Instruments INC.). Three repeats were collected along different regions of each film. The average measured thickness was 290 nm, close to the 300 nm expected from sputtering rate calculations. The electrodes after dicing were 7x14 mm. Through the use of PVC tape a 7x7 mm geometric area was defined. Lastly, the copper deposition on the electrical contact was removed by the use of 67\% \ce{HNO_3}, thus exposing the Pt underneath. This allowed to avoid undesired fluctuations in potential due to copper oxide/hydroxide formation by accidental contact with the highly alkaline electrolytic solution. Before each anodization the electrode was polarized at -1 V for 50s, to reduce the natural passivation layer on top of the exposed copper.

\subsection{Electrochemical characterization}
All electrochemical measurements were made with a Palmsens multichannel potentiostat in a 3-electrode configuration. The reference electrode was a triple junction Ag$|$AgCl$|$KCl electrode (Sigma Aldrich) with a bridge filled with 3.8M KCl solution. The bridge was emptied and refilled between each measurement. The counter electrode was a Pt coil (Sigma Aldrich). The working electrode was 7 mm x 7 mm Cu/Pt/glass electrode and the electrical connection to the plate was achieved with a tantalum clip (Redoxme) on the exposed Pt. The experiments were conducted in air at ambient conditions under a fumehood. The cyclic voltammetry and chronoamperometry experiments were performed in 50mL solutions in borosilicate glass beakers. For all of the cyclic voltammetry experiments the third cycle was taken, and unless otherwise specified the scan rate used was 100 mV/s.  All electrochemical calibrations were done in triplicate. 
To evaluate the reproducibility of our process, we replicated three times the anodizations at -0.2 V,-0.1 V and 0 V in 1 M KOH. The chronoamperometric curves were very similar, with only slight differences in current densities (\textbf{S \ref{fig:anodization-reprod}}). This difference may be attributed to the non-uniform thickness of the RF sputtering system\cite{Swann_1988} and the minimal potential instabilities of the reference electrode.
\subsection{Morphological characterization}
The morphological characterization of the samples was done with a Helios Nanolab G3 CX DualBeam FIB-SEM. Copper tape was used to ground the conductive topside of the electrode with the SEM sample holder, thus avoiding unwanted sample charging.
\subsection{X-ray photoelectron spectroscopy}
XPS was employed to investigate the surface chemical composition of the as-grown and the polarized copper films. XPS was performed in a FlexMod SPECS system equipped with a monochromatic 100W Al source (E = 1486.7 eV). The base pressure of the SPECS system is in the 10-10 mbar range. XPS spectra were recorded with a step of 0.05 eV and a pass energy of 20 eV. The XPS data were analyzed with the CasaXPS software65, and a Shirley function was used to subtract the background radiation. Spectra are energy-calibrated relative to the C1s peak of adventitious carbon (284.8 eV).
\subsection{X-ray diffraction}
The grazing incidence X-ray diffraction (GIXRD) and X-ray reflectrometry (XRR) analysis was performed with PANalytical X’Pert Pro diffractometer with a Cu K$\alpha$ source.  All GIXRD patterns are acquired without a monochromator and with an incident angle of $0.5^{\circ}$.
\subsection{Chemicals}
All chemicals were purchased from Sigma Aldrich. Glucose had a purity of $>$99.5\% and KOH pellets (reagent grade) were used to make all the alkaline solutions. L-ascorbic acid and acetaminophen had a purity of 99\%, and the latter was kept at -8°C. All chemicals were used as received, without additional purification steps. Stock solutions were prepared fresh daily in 0.1 M KOH.
\subsection{Statistical analysis}
In the course of cyclic voltammetry and chronoamperometry experiments, the recording of current, potential, and time was undertaken. Subsequently, the current values were normalized by the geometric area of the electrode (0.49 $cm^2$) to determine the current density. No additional normalization procedures were applied. Each calibration experiment, whether \textit{via} chronoamperometry or cyclic voltammetry, had a sample size of 3. When a linear interpolation was performed, the corresponding coefficient of determination $R^2$ is provided. The data is presented as mean $\pm$ standard deviation.

\section{Results and discussion}
\subsection{Electrochemical reactions for thin Cu electrode anodization}
\begin{figure} [h!]
\centering
\includegraphics[width=\linewidth]{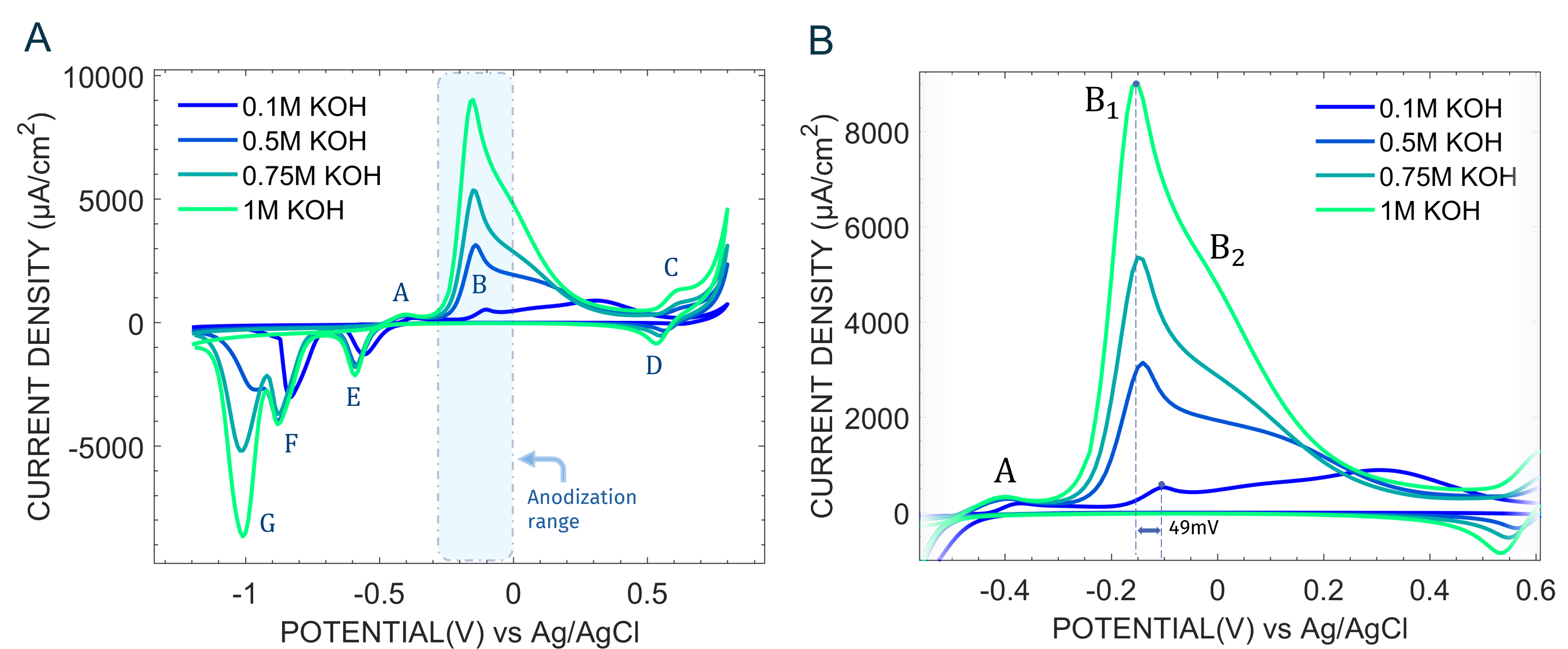}
\caption{(a) Cyclic voltammograms for Cu electrodes at different KOH concentration (0.1 M, 0.5 M, 0.75 M, 1 M), taken between -1.2V and 0.8V at a scan of 100 mV/s. (b) Zoomed-in cyclic voltammograms, highlighting the position of A, $\mathrm{B_1}$, $\mathrm{B_2}$ and the 49 mV shift going from 0.1 M KOH to 1 M KOH (here shown for $\mathrm{B_1}$) }
\label{fig:CV-cu}
\end{figure}
\begin{figure} [t]
\centering
\includegraphics[width=\linewidth]{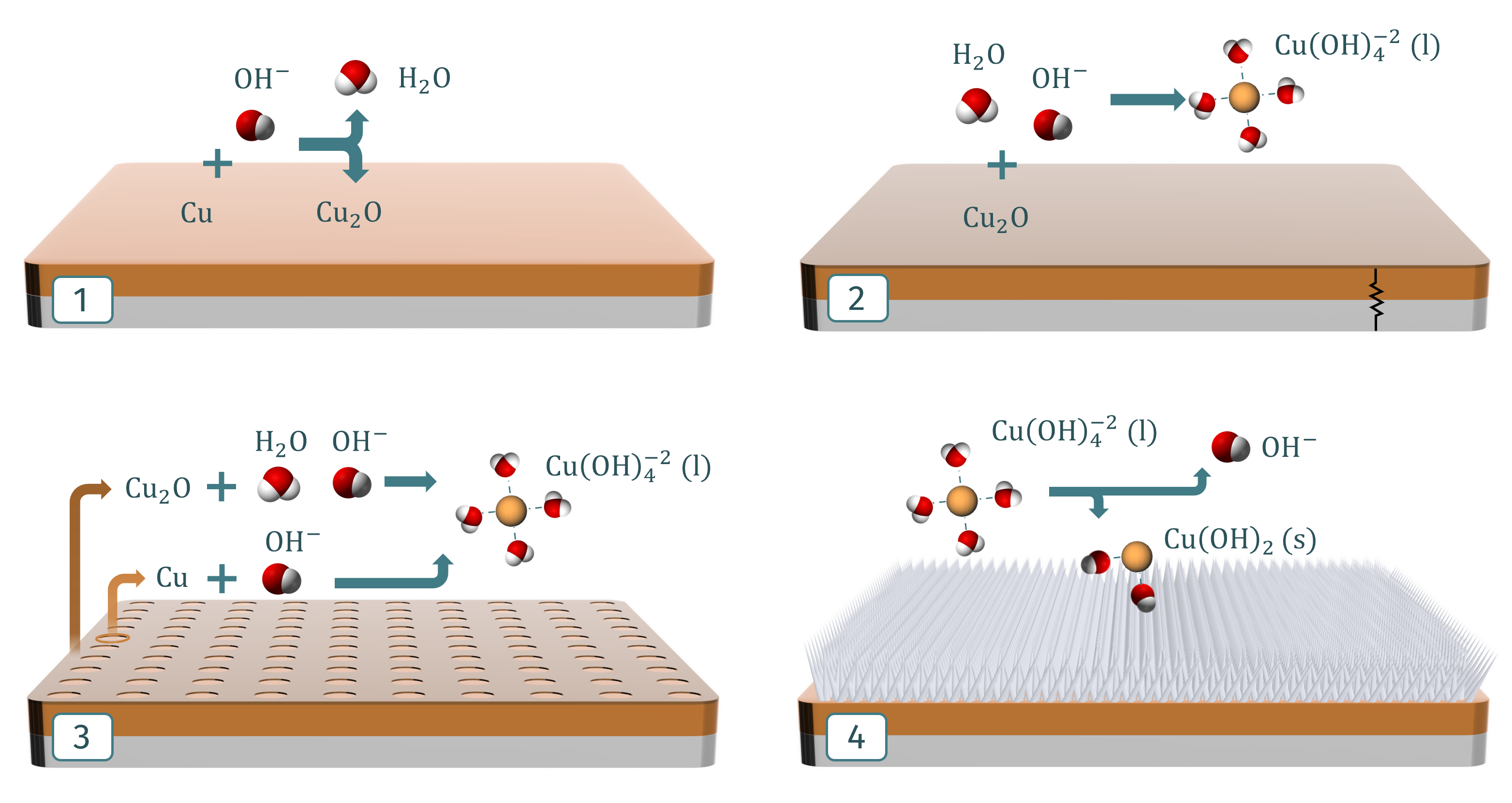}
\caption{Simplified mechanism of nanowire formation in 1 M KOH and at -0.2V. (1) Formation of insulating $\mathrm{Cu_2O}$ layer, by reaction of Cu with $\mathrm{OH^-}$. (2) Dissolution of $\mathrm{Cu_2O}$ with formation of channels which expose fresh Cu. (3) Dissolution of both $\mathrm{Cu_2O}$ and Cu. (4) The formation of $\mathrm{Cu(OH)_2}$ nanowires from supersaturated regions of $\mathrm{Cu(OH)_4^{2-}}$.  }
\label{fig:mechanismCuOx}
\end{figure}

 \noindent Prior to the selection of the anodization potential, we recorded the cyclic voltammograms of microfabricated Cu electrodes between -1.2V and 0.8V at different KOH concentrations \textbf{Figure \ref{fig:CV-cu}a}. The main features of the voltammograms match well with those reported by several researchers for polished Cu electrodes \cite{Strehblow_Maurice_Marcus_2001}. Three anodic peaks (A, B, C) and four cathodic peaks (D, E, F, G) can be discerned, associated with redox transitions of Cu(0), Cu(I), Cu(II) and - allegedly \cite{Barragan_Kogikoski_Silva_Kubota_2018} - Cu(III) for peaks C,D.  A small shoulder before A is present, although a consensus on the exact nature of this reaction is not well established \cite{Giri_Sarkar_2016}. On this, Ambrose et al.\cite{Ambrose_Barradas_Shoesmith_1973} proposed that the species $\mathrm{Cu(OH)_2^-}$ is formed by the direct electrooxidation of Cu. Based on the revised Pourbaix diagram of Cu by Beverskog et al.\cite{Beverskog_Puigdomenech_1997}, Giri et al.\cite{Giri_Sarkar_2016} later proposed that the formation of soluble $\mathrm{Cu(OH)_2^-}$ may be preceded by an intermediate step of Cu to $\mathrm{Cu_2O}$ oxidation (reaction 1 and 2). 
\begin{align}
    2 \;Cu + 2\; OH^- \rightarrow Cu_2O + H_2O + 2e^-  \label{eq:preA1}\\
    Cu_2O + 2\; OH^- + H_2O \iff 2\; Cu(OH)_2^- \label{eq:preA2}
\end{align}
As the potential is increased the rate of formation of $\mathrm{Cu_2O}$ dominates over that of dissolution, giving rise to a peak (peak A). As the potential is further increased the current increases sharply, leading to peak B (\textbf{Figure \ref{fig:CV-cu}a}). This peak can be further subdivided in two, $\mathrm{B_1}$ and $\mathrm{B_2}$ (\textbf{Figure \ref{fig:CV-cu}b}), the former occurring at more cathodic potentials. Mainly two mechanistic descriptions of the reactions underlying peak B have been advanced. One, by Haleem et al.\cite{Haleem_Ateya_1981} involves the formation of soluble or insoluble hydroxides such as $\mathrm{Cu(OH)_4^{-2}}$ and $\mathrm{Cu(OH)_2}$ respectively by the oxidation of Cu or $\mathrm{Cu_2O}$. However, this model supposes the formation of the thermodynamically unfavoured Cu(OH) and permits only one peak instead of two. Another reaction scheme was proposed by Ambrose et al. \cite{Ambrose_Barradas_Shoesmith_1973}, where the reactions underlying $\mathrm{B_1}$ and $\mathrm{B_2}$ differ depending on pH of the solution. Accordingly, peak $\mathrm{B_1}$ could be attributed to the oxidation of Cu and $\mathrm{Cu_2O}$ to soluble $\mathrm{Cu(OH)_4^{2-}}$ or insoluble $\mathrm{Cu(OH)_2}$ through reactions \ref{eq:1M-1}, \ref{eq:1M-2}, \ref{eq:0.1M}. Clearly, the greater magnitude of peak $\mathrm{B_1}$ compared to peak A suggests that reaction \ref{eq:1M-2} will dominate over \ref{eq:1M-1}, since the latter requires  $\mathrm{Cu_2O}$ formed by reaction \ref{eq:preA1}.
\begin{itemize}
    \item At 1 M KOH: 
    \begin{align}
        Cu_2O + 6\;OH^- + H_2O \rightarrow 2\; Cu(OH)_4^{2-} + 2e^-\label{eq:1M-1}\\
        Cu +4\;OH^- \rightarrow Cu(OH)_4^{2-} + 2e^- \label{eq:1M-2}
    \end{align}
    \item At 0.1 M KOH 
    \begin{equation}
        Cu_2O + 2\;OH^- + H_2O \rightarrow 2 \;Cu(OH)_2 + 2e^- \label{eq:0.1M}
    \end{equation}
\end{itemize}
Given enough time to reach supersaturation, the build up of $\mathrm{Cu(OH)_4^{2-}}$ near the surface produced by reaction \ref{eq:1M-1} and \ref{eq:1M-2} can lead to the homogeneous/heterogeneous nucleation of $\mathrm{Cu(OH)_2}$ microneedles (reaction \ref{eq:B2-1}). However, as the potential is increased close to peak $\mathrm{B_2}$ passivation reactions start to compete with the dissolution process (reaction \ref{eq:1M-1} and \ref{eq:1M-2}) that feeds reaction \ref{eq:B2-1}. Specifically, the oxidation of metallic copper by $\mathrm{OH^-}$ ions, giving $\mathrm{Cu(OH)_2}$ and CuO, as per reaction \ref{eq:B2-2} and \ref{eq:B2-3} respectively.
\begin{align}
    Cu(OH)_4^{2-}  \iff Cu(OH)_2 + 2\; OH^-  \label{eq:B2-1}\\
    Cu + 2 \; OH^- \rightarrow Cu(OH)_2 + 2e^-  \label{eq:B2-2}\\
    Cu + 2 \; OH^- \rightarrow CuO + H_2O + 2e^-   \label{eq:B2-3}
\end{align}
As a result, the chemical species on the electrode's surface in proximity to the B peaks are likely to be $\mathrm{Cu(OH)_2}$ closer to $\mathrm{B_1}$ and CuO closer to $\mathrm{B_2}$ \cite{Giri_Sarkar_2016}. Therefore, by applying a potential between the end of peak A and peak B it is possible to study the gradual change from $\mathrm{Cu_2O}$, to $\mathrm{Cu(OH)_2}$ and CuO. The presence of a potential shift of 59 mV between 1 M KOH and 0.1 M KOH for $\mathrm{B_1}$ (\textbf{Figure \ref{fig:CV-cu}b}) is particularly relevant for copper anodization research, given that significantly different species are formed in a range of approximately 300 mV. Furthermore, as previously mentioned, the possible oxidative pathways are also affected by the solution's pH, thus requiring careful consideration.\\ 
\begin{figure} [h!]
\centering
\includegraphics[width=\linewidth]{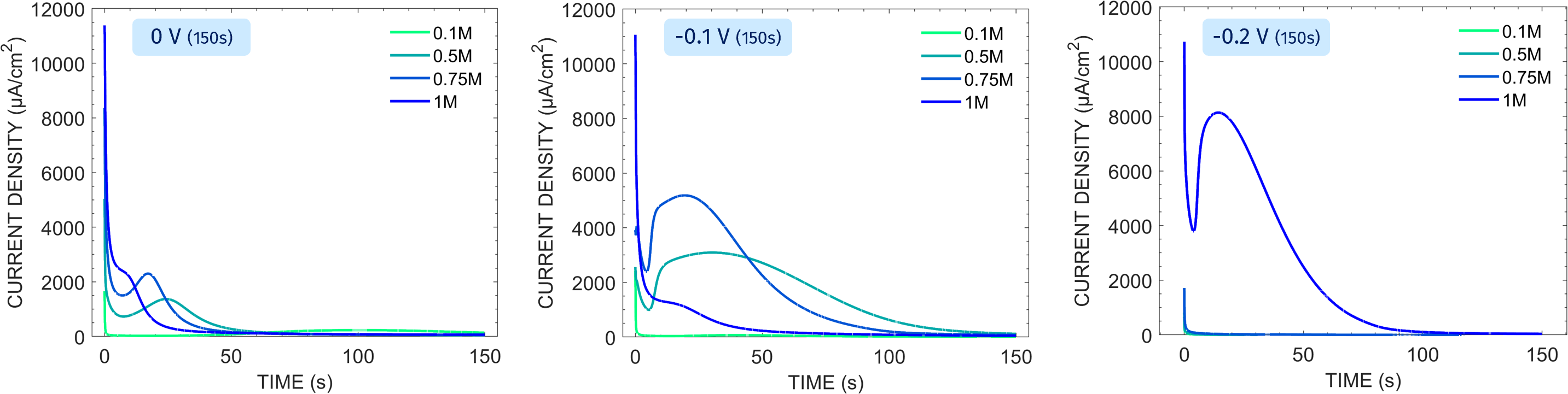}
\caption{Chronoamperometric curves at 0 V, -0.1 V, -0.2 V (from left to right) at different concentrations of KOH, each performed for 150s.}
\label{fig:anodization-KOHconc}
\end{figure} \\
\textbf{Figure \ref{fig:anodization-KOHconc}} shows the chronoamperometric curves, each lasting 150s, for microfabricated Cu electrodes obtained by applying 0V, -0.1V and -0.2V at different KOH concentrations. As illustrated, a change of 100mV in anodization potential can lead to significantly different chronoamperometric curves. For an applied potential of -0.2V and KOH concentration lower than 1 M, the current density drops rapidly due to the formation of $\mathrm{Cu_2O}$, usually of no more than 6 nm\cite{Kunze_Maurice_Klein_Strehblow_Marcus_2004,Miller_1969}.
 This process of Cu(I) oxide layer formation can be described by reaction \ref{eq:preA1}. Instead, at -0.2V for 1 M KOH concentration a completely different current time transient is obtained. The curve is characterized by an initial current drop until an inflection point is reached, then the current rises again and after reaching a maximum it slowly decays close to zero. This particular current-time response for copper has been traditionally attributed to a process of nucleation and growth of $\mathrm{Cu(OH)_2}$/CuO nanowires \cite{Stepniowski2019,Shoesmith_Rummery_Owen_Lee_1976,Becerra_Salvarezza_Arvia_1988}, in analogy with aluminum anodization \cite{Parkhutik_Shershulsky_1992}. Based on the general mechanism proposed by Shoesmith and Ambrose \cite{Shoesmith_Rummery_Owen_Lee_1976,Shoesmith_Sunder_Bailey_Wallace_Stanchell_1983,Ambrose_Barradas_Shoesmith_1973}, we can explain the chronoamperometric curve at -0.2V in 1 M KOH (\textbf{Figure \ref{fig:anodization-KOHconc}}) as follows: (i) Current density decay to a local minimum at around 4 s due to $\mathrm{Cu_2O}$ formation \cite{Shoesmith_Sunder_Bailey_Wallace_Stanchell_1983}, as per reaction \ref{eq:preA1}. The insulating layer decreases the charge transport between the electrode and the electrolyte. (ii) Dissolution of the insulating film by conversion into $\mathrm{Cu(OH)_4^{2-}}$ (reaction \ref{eq:1M-1}) with the formation of channels in the oxide that expose fresh Cu; (iii) Dissolution of Cu by reaction \ref{eq:1M-2} with the formation of $\mathrm{Cu(OH)_4^{2-}}$. (iv) Formation high $\mathrm{Cu(OH)_4^{2-}}$ concentration regions, that convert into $\mathrm{Cu(OH)_2}$ (reaction \ref{eq:B2-1}), growing into nanowires. The nucleation of hydrous $\mathrm{Cu(OH)_2}$ requires supersaturated $\mathrm{Cu(OH)_4^{2-}}$ near the electrode's surface, as shown by previous work where upon electrode rotation the reaction is inhibited \cite{Shoesmith_Rummery_Owen_Lee_1976} (v) The current increases until a maximum is reached at around 20 s, corresponding to the point where the $\mathrm{Cu(OH)_2}$ crystals start to impinge upon one another, thereby progressively impeding reaction \ref{eq:1M-1} and \ref{eq:1M-2}. As a result, the current gradually falls close to zero, with nanowire growth halting at t=120s. A simplified illustration of the mechanism of nanowire formation is shown in \textbf{Figure \ref{fig:mechanismCuOx}}.
The change in nanowire morphology at different points in the current-time curve for an anodization at -0.2V (vs $\mathrm{Ag|AgCl_{sat}}$) and 1 M KOH has been studied by  Stepniowski et al. \cite{Stepniowski2019}. The authors noted a progressive increase in nanowire thickness and length for longer anodization times.
On the other hand, the anodizations at 0 V (\textbf{Figure \ref{fig:anodization-KOHconc}}) show dramatically lower current densities while still maintaining the same nucleation and growth profile. This can be explained by noting that 0V in 0.5 M, 0.75 M and 1 M KOH is closer to $\mathrm{B_2}$ than $\mathrm{B_1}$.  Hence, at polarization the Cu surface transforms into $\mathrm{Cu_2O}$ (reaction \ref{eq:preA1}) but also into CuO (reaction \ref{eq:B2-3}). This limits the formation of $\mathrm{Cu(OH)_4^{2-}}$, as reaction \ref{eq:1M-1} and \ref{eq:1M-2} can happen only on the available Cu or $\mathrm{Cu_2O}$ sites. As a result, under these conditions the predominant surface species will be CuO, instead of  $\mathrm{Cu(OH)_2}$ or $\mathrm{Cu_2O}$. In general, all anodizations done in 0.1 M KOH show a sharp decline current densities independently of the applied potential. This is because, as reported by Ambrose et al. \cite{Ambrose_Barradas_Shoesmith_1973}, reaction \ref{eq:1M-2} is replaced by reaction \ref{eq:preA1} and \ref{eq:0.1M}. Therefore, the process of dissolution is severely limited, likely resulting in the formation of a $\mathrm{Cu_2O}$ layer. Lastly, at an applied potential of -0.1 V we can observe (\textbf{Figure \ref{fig:anodization-KOHconc}}) that the current-time transients are also typical of a nucleation and growth process (except in 0.1 M KOH). Here, an intermediate behavior can be noted between the anodizations at 0V and -0.2V, where at -0.1V in 0.5 M KOH the curve resembles that at -0.2V and 1 M KOH, and that at -0.1V in 1 M KOH resembles that at 0V in 0.5 M KOH. This change the shape of the curves at -0.1V, associated with a reduction of current density for increasing KOH concentration, can be attributed to the progressively more favourable formation of CuO which impedes further etching and thus competes with the process of nanowire formation\cite{Becerra_Salvarezza_Arvia_1988}.
\clearpage

\subsection{Electrodes' morphological and compositional characterization}
After anodizing \textit{via} chronoamperometry the copper electrodes under different potentials (-0.2 V, -0.1 V, 0 V) and KOH molarities (0.1 M, 0.5 M, 0.75 M, 1 M), we selected 4 configurations whose chemical and morphological state was expected to differ significantly. Specifically: -0.1 V in 0.5 M, -0.2 V in 0.75 M, 0 V in 1 M and a bare Cu electrode. To simplify the nomenclature we will henceforth refer to them by (S+applied potential), with the bare referred to as SBare. All the morphological, chemical and crystallographic characterizations were performed after 3 cycles in 0.1 M KOH ranging from 0 V to 0.8 V. This approach allowed us to consider the stable state conditions of use.  
\begin{figure} [h!]
\centering
\includegraphics[width=\linewidth]{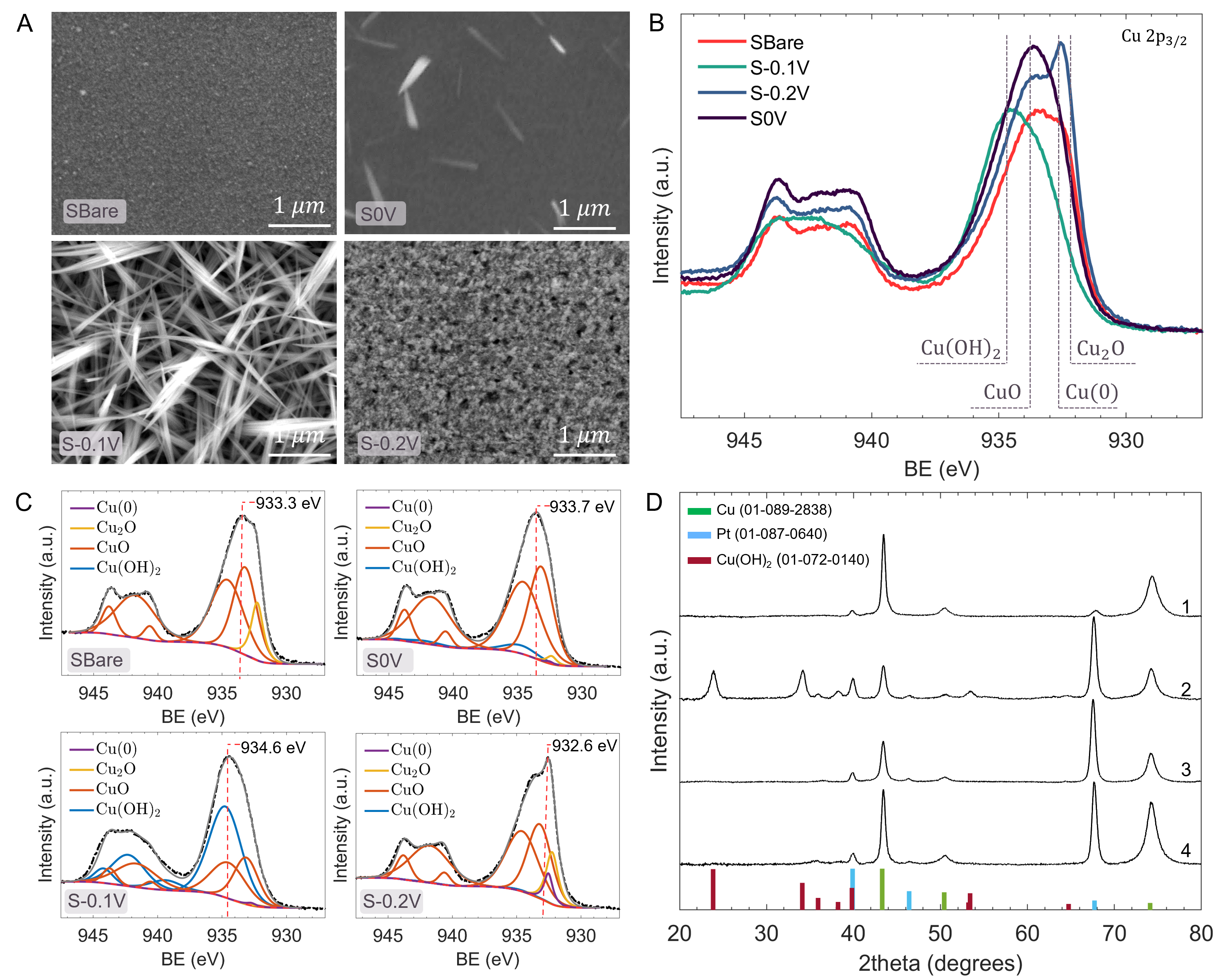}
\caption{(A) SEM images of SBare, S0V, S-0.1V and S-0.2V.(B) Cu $\mathrm{2p_{3/2}}$ XPS spectra combined, overlayed with reference binding energies (C) Cu $2p_{3/2}$ XPS spectra of SBare, S0V, S-0.1V, S-0.2V samples with corresponding fit. (D) XRD diffraction patterns for SBare (D.1), S-0.1V (D.2), S-0.2V (D.3) and S0V (D.4).}
\label{fig:SEMXPS}
\end{figure}\\

Scanning electron microscopy (SEM) was conducted to examine the surface morphology, as illustrated in (\textbf{Figure \ref{fig:SEMXPS}a}). SBare exhibits a flat surface, devoid of significant features, suggesting that cycling between 0V and 0.8V promotes the formation of a passivating film, which effectively shields the surface from corrosion. Similarly, S0V maintains the predominantly flat morphology observed in SBare, albeit with the presence of short, sparse nanowires. In stark contrast, S-0.1V displays a dense network of nanowires, each with a diameter of approximately 10 nm, and these nanowires have formed in clusters with an average cluster diameter of $122 \pm 21$ nm. A detailed comparison between S-0.1V and SBare at higher magnification (50kX) is provided in \textbf{S\ref{fig:SEMzoom}}. Finally, S-0.2V is distinguished by a porous structure, likely resulting from a competitive interplay between reactions \ref{eq:preA1} and \ref{eq:preA2}. The samples were analyzed using X-ray photoelectron spectroscopy (XPS) to evaluate their surface chemical states. A survey spectrum of SBare (\textbf{S \ref{fig:surveyXPS}}) reveals prominent core-level lines associated with copper, including a Cu 2p doublet (Cu $\mathrm{2p_{1/2}}$ and $\mathrm{2p_{3/2}}$) at 930-955 eV, Cu 3s at 123 eV, Cu 3p at 76 eV, and Cu 3d at 3 eV. Peaks corresponding to oxygen (O 1s) at 530 eV and adventitious carbon (C 1s) at 285 eV are also present. \textbf{Figure \ref{fig:SEMXPS}b} shows the Cu $\mathrm{2p_{3/2}}$ signals for electrodes biased at various potentials, with the binding energies of relevant copper species superimposed. By comparing the main emission band's peak positions with literature values for Cu(0), $\mathrm{Cu_2O}$, CuO, and $\mathrm{Cu(OH)_2}$, qualitative insights can be drawn. All samples display a well-defined shakeup satellite, indicating the presence of Cu(II) species\cite{Otamiri_Andersson_Andersson_1990,Poulston_Parlett_Stone_Bowker_1996,Salvador_Fierro_Amador_Cascales_Rasines_1989,Wang_Miller_Notis_1996}. The main peak of SBare is at a binding energy of 933.3 eV, close to that of pure CuO at $933.8 \pm 0.2$ eV \cite{Kautek_Gordon_1990}. Additionally, a small shoulder at lower binding energies corresponds to either metallic Cu or $\mathrm{Cu_2O}$, found at $932.6 \pm 0.2$ eV and $932.4 \pm 0.2$ eV, respectively \cite{Kautek_Gordon_1990}. This shoulder disappears in S0V, with the peak maximum shifting to 933.7 eV, suggesting an increase in the CuO component at the expense of Cu(0) and $\mathrm{Cu_2O}$. The peak maximum in the $\mathrm{L_{3}M_{4,5}M_{4,5}}$ Auger spectrum for S0V is at a kinetic energy of 917.5 eV, close to the reference value of 917.6 eV for CuO \cite{Biesinger_2017}, and slightly lower than SBare at 917.9 eV. S-0.1V shows a peak maximum at 934.6 eV, closely matching $\mathrm{Cu(OH)_2}$ at $934.5 \pm 0.2$ eV \cite{Kautek_Gordon_1990}. A similar shift is seen in the $\mathrm{L_{3}M_{4,5}M_{4,5}}$ Auger spectrum, with its maximum at a kinetic energy of 916.4 eV, near 916.25 eV for pure $\mathrm{Cu(OH)_2}$. Moreover, the shakeup feature in S-0.1V differs from all other samples, exhibiting the typical shape of $\mathrm{Cu(OH)_2}$ reported in the literature\cite{Biesinger_2017}. As expected, the O1s spectrum of S-0.1V shows a dominant Cu-OH component at a binding energy of 531.1 eV (\textbf{S \ref{fig:O1sC1s}}). In contrast, the shoulder at lower binding energy observed in SBare is more pronounced in S-0.2V, shifting the peak maximum to 932.6 eV. This suggests that the polarization at -0.2V increases the Cu(0) or $\mathrm{Cu_2O}$ fraction at the expense of CuO, consistent with thermodynamic predictions \cite{Parsons_1967}. Despite the increased Cu(0) or $\mathrm{Cu_2O}$ fraction, the presence of a dominant CuO component means that the $\mathrm{L_{3}M_{4,5}M_{4,5}}$ Auger spectrum peak position does not significantly shift from SBare to S-0.2V, with only a slight shoulder emerging at higher kinetic energies. Quantifying the relative contributions of different species in mixed-oxidation-state copper electrodes is challenging due to significant peak overlap. To address this, the well-established fitting procedure developed by Biesinger et al. \cite{Biesinger_2017} was employed. This method fits the entire Cu $\mathrm{2p_{3/2}}$ peak shape using a set of peaks for each species, constrained in binding energy, full width at half maximum (FWHM), and relative areas. Peak positions and relative areas were referenced to the first peak of each species, with variations of $\pm 0.1$ eV allowed to account for the imprecision due to the C1s calibration. Similarly, the FWHM was allowed to vary by $\pm 0.1$ eV. This approach, while similar to using actual line shapes from pure compounds, offers greater flexibility in its application. \textbf{Figure \ref{fig:SEMXPS}c} shows the fitted Cu $\mathrm{2p_{3/2}}$ spectra of SBare, S0V, S-0.1V, and S-0.2V. The fitting parameters, along with the fractional areas of each component, are detailed in the supplementary information (\textbf{S Table \ref{tab:XPS-SBare}, S Table \ref{tab:XPS-S0V}, S Table \ref{tab:XPS-S-0.1V}, S Table \ref{tab:XPS-S-0.2V}}). Consistent with the previous qualitative analysis, SBare is primarily composed of CuO (89\%), followed by Cu(0) + $\mathrm{Cu_2O}$ (11\%). In S0V, the CuO fraction increases to 93\%, with some $\mathrm{Cu(OH)_2}$ (5\%) and only trace amounts of Cu(0) + $\mathrm{Cu_2O}$ (1\%). S-0.1V is dominated by $\mathrm{Cu(OH)_2}$ (55\%), followed by CuO (45\%). S-0.2V, compared to SBare, has a lower CuO fraction (84\%) but an increased Cu(0) + $\mathrm{Cu_2O}$ component (15\%).\\
X-ray diffraction (XRD) analysis was also conducted to gain additional insights, with the corresponding diffractograms shown in \textbf{Figure \ref{fig:SEMXPS}d}. Regardless of polarization potential, the anodization does not appear to affect the entire volume of the thin copper film, as evidenced by the presence of diffraction peaks for Cu and Pt in all samples (SBare, S0V, S-0.2V, S-0.1V). For SBare, S0V, and S-0.2V, the diffraction pattern is dominated by the Pt/Cu bilayer signal, suggesting that the thickness of the produced anodic oxides is too small to be reliably detected by XRD. Copper diffraction peaks are observed at $2\theta$ values of 43.3, 50.4, and 74.1 degrees, with corresponding Miller indices (\textit{hkl}) of (111), (200), and (220). Platinum diffraction peaks are noted at $2\theta$ values of 39.9, 46.4, and 67.4 degrees, with corresponding Miller indices of (111), (200), and (220). These planes indicate that a cubic phase structure can be assigned to both the copper (ICSD01-089-2838) and platinum (ICSD01-087-0640) layers. In contrast, the XRD pattern of S-0.1V indicates that the produced nanowires are crystalline and primarily composed of $\mathrm{Cu(OH)_2}$. Diffraction peaks of $\mathrm{Cu(OH)_2}$ are observed at $2\theta$ values of 23.8, 34.1, 35.9, 38.2, 39.8, 53.2, 53.4 and 64.6 degrees, with corresponding Miller indices of (021), (002), (111), (041), (022), (130), (150), (132), and (152). These planes suggest an orthorhombic crystal structure for $\mathrm{Cu(OH)_2}$ (ICSD01-072-0140). It is possible that some peaks associated with CuO may also be present, but they might be too weak to be distinguished from adjacent $\mathrm{Cu(OH)_2}$ peaks. The average crystallite size was estimated using the Scherrer formula \cite{scherrer1918nachr}.

\begin{equation}
    \mathrm{d=\frac{0.9 \lambda}{\beta \cdot cos\theta}}
\end{equation}
Where ‘$\lambda$’ is wave length of X-Ray (0.1541 nm), ‘$\beta$’ is FWHM (full width at half maximum),
‘$\theta$’ is the diffraction angle and ‘d’ is the mean crystallite diameter. To estimate the crystallite size for the $\mathrm{Cu(OH)_2}$ nanowires of S-0.1V we considered the three most intense peaks ($2\theta=23.8^\circ$, $34.1^\circ$, $53.4^\circ$) and obtained a value of $12 \pm 2$ nm. For the Cu and Pt layers on SBare, the three most intense peaks were considered for each. For Pt ($2\theta=39.9^\circ$, $46.4^\circ$, $67.4^\circ$) the mean crystallite size was $14\pm 2$ nm; For Cu ($2\theta=43.3^\circ$, $50.4^\circ$, $74.1^\circ$) it was $11 \pm 4$ nm.

\subsection{Effect of process parameters on glucose oxidation at anodized Cu electrodes}
Once the microfabricated electrodes were anodized under different potentials (-0.2V, -0.1V, 0V) and KOH concentrations (0.1 M, 0.5 M, 0.75 M, 1 M), we proceeded to evaluate their electrocatalytic performance towards glucose electrooxidation.  The  selection of a potentiodynamic method, such as cyclic voltammetry, over a potentiostatic one aimed at providing additional metrics to assess the oxidation process. Of these, the shape of the CVs and the current density at different potentials can point to different mechanisms\cite{Elgrishi_Rountree_McCarthy_Rountree_Eisenhart_Dempsey_2018,Bard_Faulkner}. Thus, we acquired the CVs of the bare and the anodized electrodes in a solution of 0.1 M KOH with 8 mM of glucose.
\begin{figure} [h!]
\centering
\includegraphics[width=\linewidth]{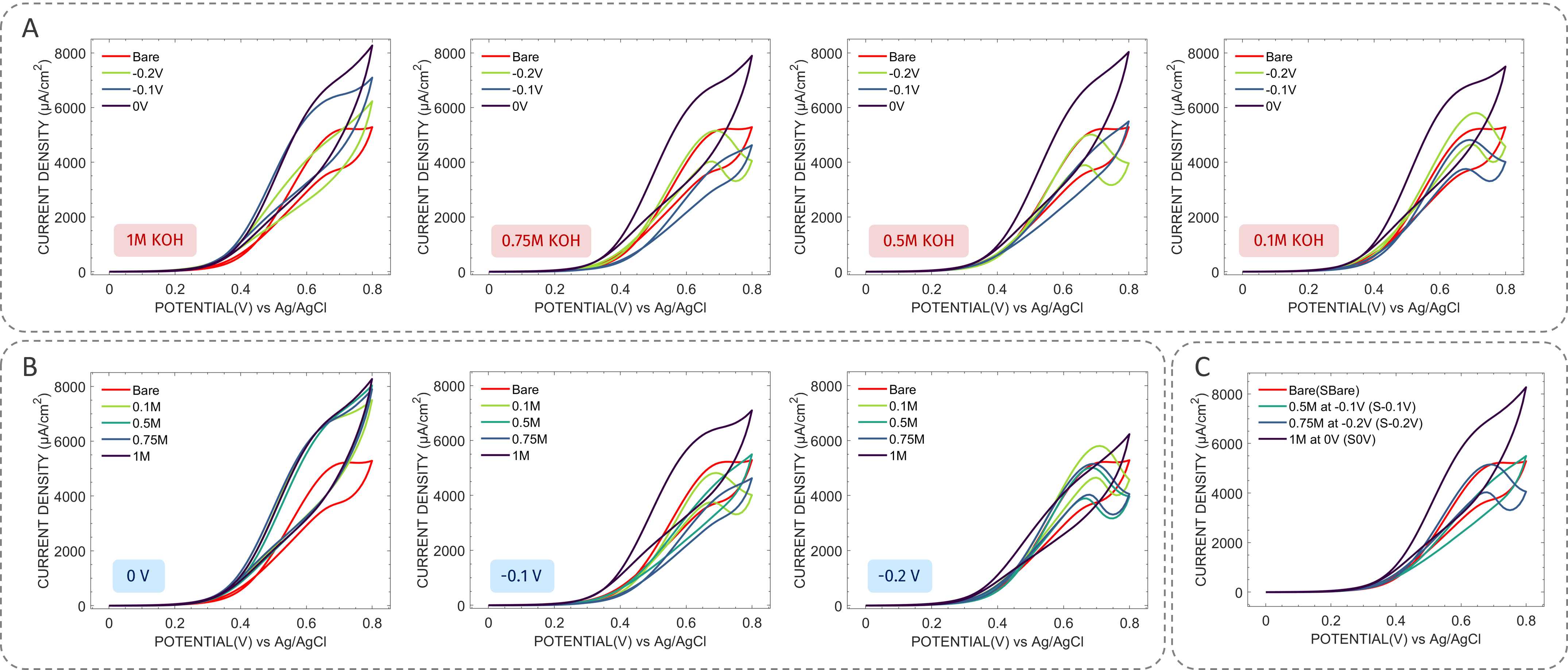}
\caption{Cyclic voltammograms in 0.1 M KOH and 8 mM glucose at 100 mV/s scan rate. (A) Effect of the anodization potential for a fixed concentration of KOH. (B) Effect of the KOH concentration for a fixed anodization  potential. (C) Cyclic voltammograms of the samples selected for morphological, chemical and crystallographic analysis.}
\label{fig:CV8mMglucose}
\end{figure}\\
\noindent
\textbf{Figure \ref{fig:CV8mMglucose} a,b} compare respectively the obtained CVs at the third cycle as a function of the KOH molarity of the anodization solution and of the applied anodization potential. Here we show that anodizations done on a reduced copper surface introduce stable chemical and morphological changes that can either favour and disfavour glucose oxidation. The CV of the bare Cu in a glucose solution (red curve in \textbf{Figure \ref{fig:CV8mMglucose}c}) presents the typical plateau profile which has been traditionally ascribed to a Cu(II) to Cu(III) transition \cite{Aun_Salleh_Ali_Manan_2021}. Additionally, the symmetric shape of the CV \cite{Ostervold_Bakovic_Hestekin_Greenlee_2021}, combined with the decreasing sensitivity from cycle to cycle (\textbf{S \ref{fig:sensitivity-scan}}) would suggest the presence of an adsorption step of oxidation products. Xie et al. \cite{Xie_Huber_1991} advanced the existence of two limiting mechanisms for glucose oxidation on copper electrodes: on the left side of the peak the hydroxide adsorption is limiting, whereas on the right side the adsorption of glucose is limiting. Inspired by the study of  Fleischmann et al.\cite{Fleischmann1972TheKA} on nickel oxide, the authors proposed the adsorbed hydroxyl radical to be the main oxidizer, with the rate determining step being the H abstraction from the $C_\alpha$ and the formation of a bridged cyclic intermediate. In their model, as the electrode is polarized anodically, $\mathrm{Cu_2O}$ would tranform to CuO according to the following reaction:
\begin{equation}
    Cu_2O+2OH^-\rightarrow 2CuO + H_2O + 2e^-
    \label{eq:Cu2otoCuO}
\end{equation}
Subsequently, Cu(III) species would form and mediate the oxidation of glucose. However, this is still a contested issue, given that thermodynamic predictions indicate its formation potential to be more anodic than OER.  Lately, Barragan et al. \cite{Barragan_Kogikoski_Silva_Kubota_2018} further put into question the formation of Cu(III) species, proposing instead that a partial charge transfer occurs between the vacancies in the CuO and the adsorbed hydroxide ions. Reaction \ref{eq:pairingh+} represents the coupling of the vacancies $\mathrm{h^+}$ and the adsorbed OH, and reaction \ref{eq:chargetransf} shows the formation of OH radical through effective charge transfer. The latter is expected to happen to a lesser extent, according to the authors \cite{Barragan_Kogikoski_Silva_Kubota_2018}.
\begin{align}
OH^-_{ads} + h^+ \xrightarrow[]{k(E)}& (OH^-_{ads})(h^+) \label{eq:pairingh+} \\ 
(OH^-_{ads})(h^+) \xrightarrow[]{k(E)}& OH^\cdot_{ads} \label{eq:chargetransf} \\ 
(OH^-_{ads})(h^+) \; + \; RCH_3OH \rightarrow & H_2O_ {ads} + RCH_2OH \label{eq:glucoseox}
\end{align}
The energy accumulated in the pairing between the vacancies $\mathrm{h^+}$ and $\mathrm{OH^-}$ catalyses the oxidation of glucose through reaction \ref{eq:glucoseox}. Through a two electron transfer, gluconolactone and two water molecules are expected to form, succeeded by additional oxidation reactions.
Within this framework, the initial step of glucose oxidation doesn't require the formation of a Cu(III)-glucose complex but leverages the semiconductive properties of CuO. This would imply that CuO formation is a necessary step to produce the active $\mathrm{(OH_{ads}^-)(h^+)}$ sites. Our results support this model: for 1 M KOH we observe a progressive increase in the glucose oxidation current for increasing anodization potential (\textbf{Figure \ref{fig:CV8mMglucose}a}). At the same time, the peak current (excluded the component from oxygen evolution) shifts cathodically, indicative of a decreasing energetic barrier to glucose electrooxidation. This is due to a greater fraction of CuO to $\mathrm{Cu(OH)_2}$, as the anodization is performed at potentials gradually closer to  $\mathrm{B_2}$ and further to $\mathrm{B_1}$ (\textbf{Figure \ref{fig:CV-cu}}). Similarly, for a 0V polarization the KOH molarity has a slight positive effect on the glucose oxidation current density - albeit not as dramatic as the applied potential. This may be due to a 59 mV cathodic shift going from 0.1 M to 1 M, as previously mentioned. Consequently, the the most performing electrode is obtained by polarizing at 0V in 1 M KOH (S0V), as it maximizes the fraction of the catalytically active cupric oxide. \\ The CVs at -0.1V for varied KOH concentrations (\textbf{Figure \ref{fig:CV8mMglucose}b}) show very distinct profiles. As this potential is at the cross point of A,$\mathrm{B_1}$,$\mathrm{B_2}$, the description of all the represented CVs allows to cover all the observed variability within the [KOH molarity-potential] space.  At -0.1V for 0.1 M of KOH, peak A is favoured and the current densities are similar to that of the bare Cu (SBare) until they drop at 0.6V. Its shape strongly resembles that of sample S-0.2V (\textbf{Figure \ref{fig:CV8mMglucose}c}), which as expected presented a greater $\mathrm{Cu_2O}$ fraction (\textbf{S Table \ref{tab:XPS-S-0.2V}}). The reduced currents at $>0.6$V may be due to a lower affinity for oxygen evolution or for successive glucose oxidation oxidation steps.  Although $\mathrm{Cu_2O}$ is expected to convert to CuO through reaction \ref{eq:Cu2otoCuO}, the quick scanning of potentials in a CV is likely to affect only a superficial layer. Hence, the amount of $\mathrm{(OH_{ads}^-)(h^+)}$ is limited by the extent of $\mathrm{Cu_2O}$ being transformed into CuO. Conversely, at  M KOH the voltammogram rises more slowly than the bare Cu, until reaching the same current densities at 0.8V. Under these conditions $\mathrm{B_1}$ is favoured and $\mathrm{Cu(OH)_2}$/CuO nanowires form, as highlighted by our characterization of sample S-0.1V (\textbf{Figure \ref{fig:SEMXPS}}). The presence of $\mathrm{Cu(OH)_2}$ nanowires appears to introduce an overpotential, slowing down the electrooxidation process. Barragan et al. \cite{Barragan_Kogikoski_Silva_Kubota_2018}, reported for $\mathrm{Cu(OH)_2}$ an intermediate activity between Cu and CuO. However, our findings indicate the copper hydroxide sample (S-0.1V) to be the least performing. This could be attributed to a morphological difference, as our samples had a dense layer of $\mathrm{Cu(OH)_2}$ nanowires, instead of a film obtained through chemical means. Nonetheless, improved catalytic activity was reported going from $\mathrm{Cu(OH)_2}$ to CuO.
S. Zhou et al.\cite{Zhou_Feng_Shi_Chen_Zhang_Song_2013} in their Cu anodization work found $\mathrm{Cu(OH)_2}$ to be more active for glucose sensing than CuO. Nevertheless, the study lacked a comparison with a Cu electrode, and their CuO electrode may have under performed due to a significantly longer anodization process, resulting in a thicker oxide layer, as compared to their $\mathrm{Cu(OH)_2}$ electrode. 
 \begin{figure} [h!]
\centering
\includegraphics[width=0.65\linewidth]{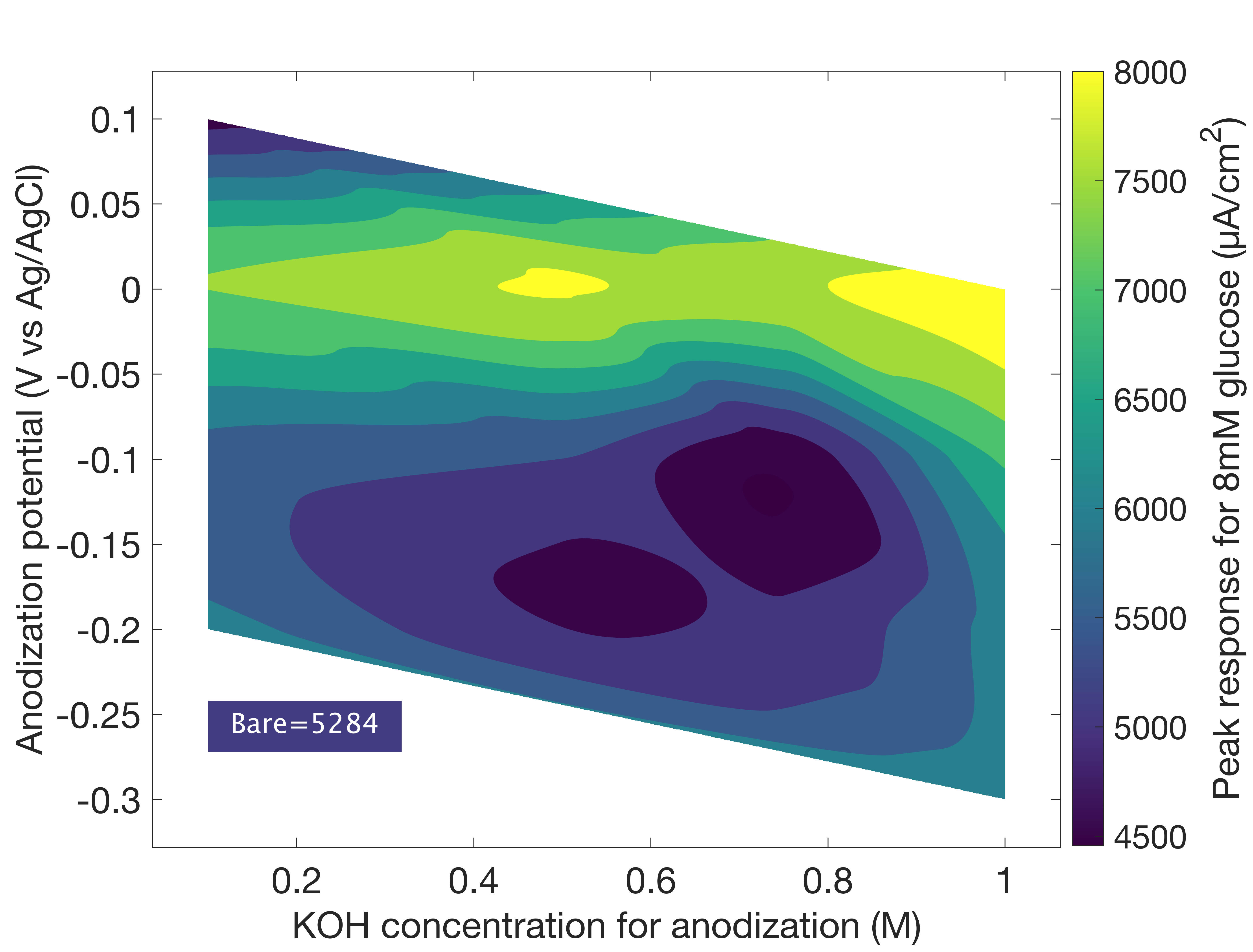}
\caption{Contour plot of the effect of process parameters on the glucose oxidation current}
\label{fig:controurplot}
\end{figure} 
\\The effect of the anodization parameters (KOH concentration and anodization potential) on the glucose oxidation current in CV is summarized by the a contour plot in \textbf{Figure \ref{fig:controurplot}}. The graph was produced by recording the peak current density of each CV at 8 mM glucose in 0.1 M KOH for a total of 15 unique combinations of anodization parameters (\textbf{S Table \ref{tab:countourdata}}). Afterwards, a cubic 2D interpolation was performed in order to transform a discrete set of data points into a surface. Accordingly, applying a potential of 0 V leads to the greatest performance improvements, with 1 M KOH being the best electrolyte concentration. Moreover, at potentials between -0.1 V and -0.2 V and KOH concentrations between 0.4 M and 0.8 M a region of impaired catalytic activity is observed, where the response is worse than for a bare Cu electrode. Within this range $\mathrm{Cu(OH)_2}$ nanowires are formed, as confirmed by our characterization of S-0.1V (\textbf{Figure \ref{fig:SEMXPS}}), the chronoamperometry curves (\textbf{Figure \ref{fig:anodization-KOHconc}}) and the clear resemblance in shape of the CVs in 8 mM glucose (\textbf{Figure \ref{fig:CV8mMglucose}}). 

\subsection{Sensing performance towards glucose \textit{via} cyclic voltammetry and chronoamperometry.}
Our systematic study on the effect of the anodization parameters on the electrocatytic activity towards glucose clearly highlighted that an anodization step at 0 V in 1 M KOH led to the greatest current density for 8 mM glucose in 0.1 M KOH (\textbf{Figure \ref{fig:controurplot}}, \textbf{S Table \ref{tab:countourdata}}). Although a single concentration CV can be used as a useful screening method, the different response towards oxygen evolution superimposing to the glucose oxidation could inflate the actual sensitivity increase. Consequently, we performed a full calibration with cyclic voltammetry and chronoamperometry of sample S0V and the bare coppper sample SBare.
\begin{figure} [ht]
\centering
\includegraphics[width=\linewidth]{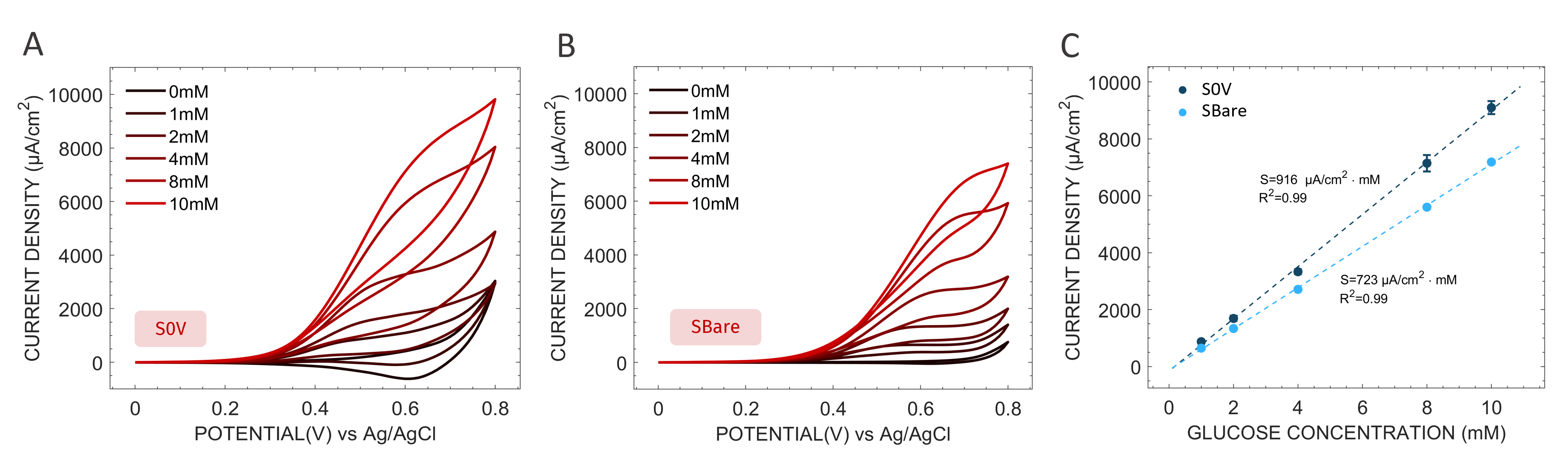}
\caption{(A) Cyclic voltammograms in 0.1 M KOH for the S0V sample with a glucose concentration between 0 mM-10 mM. (B) Cyclic voltammograms in 0.1 M KOH for the SBare sample with a glucose concentration between 0 mM-10 mM. (C) Calibration plot of the glucose oxidation peak for S0V and SBare, third cycle considered. All data is expressed as a mean ± SD (n=3). }
\label{fig:CVcal}
\end{figure}\\
\textbf{Figure \ref{fig:CVcal}} shows the CVs at the third cycle for S0V and SBare, with the corresponding calibration plot. Both samples show very good linearity in the 1-10 mM range, with S0V having a sensitivity of 916 $\pm$ 17 µA/$\mathrm{cm^2 \cdot mM}$, against a sensitivity of 723 $\pm$ 9 µA/$\mathrm{cm^2 \cdot mM}$ for SBare - equivalent  to a 26 $\%$ increase. Although it is not always specified, the cycle number has a significant effect on the sensor's response. For S0V, a loss of $\approx 25 \%$ of sensitivity is observed from the first to the third cycle, after which it stabilizes (\textbf{S \ref{fig:sensitivity-scan}}). Whereas for SBare, this sensitivity drop is only $\approx 17 \%$. As a result, the sensitivity gain between S0V and SBare reduces from $40\%$ to $26\%$ between the first and third cycle. Given that CuO and Cu electrodes are generally considered to be stable within this potential range \cite{Aun_Salleh_Ali_Manan_2021}, the observed current reduction is likely associated with the adsorption of oxidation products \cite{Ostervold_Bakovic_Hestekin_Greenlee_2021}, of which S0V appears to be more susceptible. S0V might favour additional electrooxidation steps whose products tend to temporarily block the active sites. The limit of detection (LOD) was calculated using the following formula \cite{IUPAC}:
\begin{equation}
    \mathrm{LOD}=\frac{3.3 \cdot \sigma_{\mathrm{blank}}}{S}
    \label{eq:LOD}
\end{equation}
Where $\sigma_{\mathrm{blank}}$ is the standard deviation of the blank and S is the sensitivity. Here, we obtained $\sigma_{\mathrm{blank}}$ by evaluating the current densities of three blanks at 0.46V, corresponding to the potential of the peak at the lowest concentration (1 mM). For S0V and SBare respectively, the LOD in the 1 mM-10 mM range is 1.35 mM and 0.81 mM. 
We further complemented the potentiodynamic study by recording the CVs for S0V and SBare with a glucose concentration of 2 mM in 0.1 M KOH at different scan rate (from 25 mV/s until 200 mV/s in steps of 25 mV/s). The peak glucose current was observed to be linearly dependent on the square root of the scan rate for both samples (\textbf{S \ref{fig:scanrate}}), indicative of diffusion limited process\cite{Bard_Faulkner}.  Here we considered the current density at 0.6 V to minimize the interference of oxygen evolution.
\begin{figure} [ht]
\centering
\includegraphics[width=\linewidth]{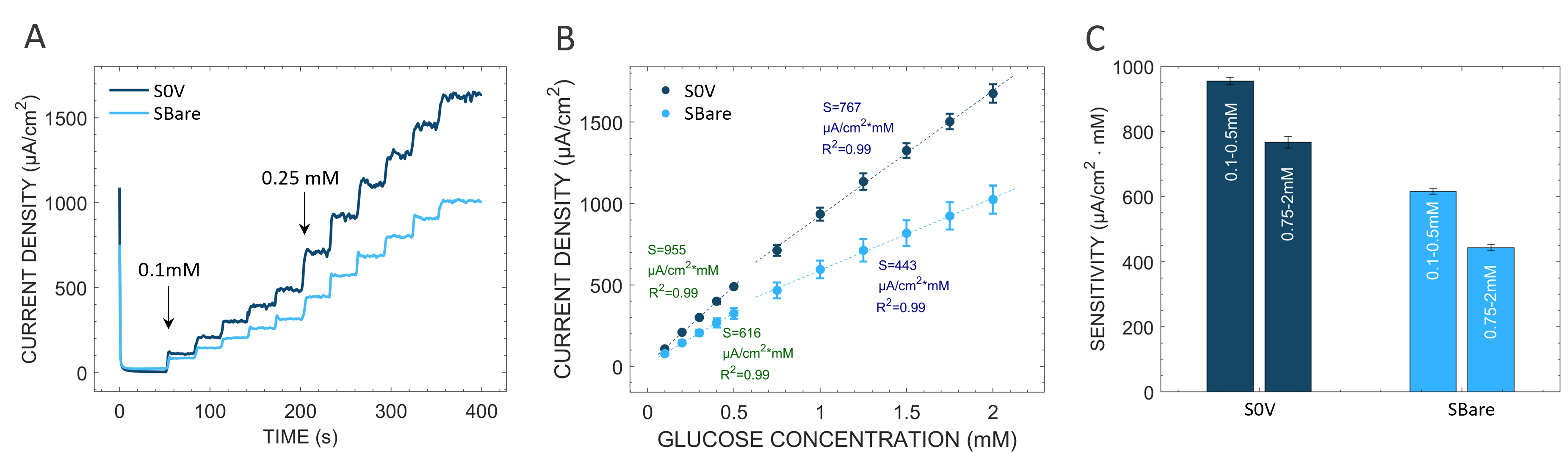}
\caption{(A) Chronoamperometric curves of S0V (dark blue) and SBare (light blue) in 0.1 M KOH at 0.5 V showing additions of 0.1 mM of glucose until 0.5 mM and then of 0.25 mM until 2 mM.(B) Calibration plot for S0V (dark blue) and SBare (light blue) for an applied potential of 0.5 V. (C) Corresponding sensitivities in the 0.1-0.5 mM and 0.75-2 mM ranges for S0V (dark blue) and SBare (light blue). All data is expressed as a mean ± SD (n=3)}
\label{fig:Chronocal}
\end{figure} 
\\
\textbf{Figure \ref{fig:Chronocal}a} presents the chronoamperometric curves for S0V and SBare with the addition of 0.1 mM of glucose until 0.5 mM and then in steps of 0.25 mM until 2 mM. \textbf{Figure \ref{fig:Chronocal}b} includes the calibration curves obtained from three independent chronoamperometry experiments each, with the calculated sensitivities in the bar plot in \textbf{Figure \ref{fig:Chronocal}c}. To obtain a calibration curve we averaged the value of the current density in the middle 10s for each 30s step, in order to minimize fluctuations. The sensitivity in the 0.1 mM to 0.5 mM range for sample S0V and SBare respectively is 955 $\pm$ 11 µA/$\mathrm{cm^2 \cdot mM}$ and 616 $\pm$ 9 µA/$\mathrm{cm^2 \cdot mM}$, equivalent to a $55\%$ gain over SBare. Whereas, in the 0.75 mM to 2 mM S0V had a sensitivity of 767 $\pm$ 18 µA/$\mathrm{cm^2 \cdot mM}$ and SBare of 443 $\pm$ 10 µA/$\mathrm{cm^2 \cdot mM}$, with a corresponding gain of $73\%$ over SBare. The LOD was calculated by using equation \ref{eq:LOD}, where $\sigma_{\mathrm{blank}}$ corresponds to the standard deviation for three samples of the mean current density from 40s to 45s after polarization, prior to the first glucose addition. For S0V and SBare respectively, the LOD in the 0.1 mM-0.5 mM range is 0.0041 mM and 0.0597 mM, while in the 0.75 mM-2 mM range is 0.0051 mM and 0.08 mM.
Consequently, cyclic voltammetry may not be the most appropriate electroanalytical technique for glucose sensing on microfabricated copper electrodes due to the deactivation process observed from the first to the second cycle, combined with its relatively high LOD. In contrast, chronoamperometry allows to obtain an order of magnitude lower LOD and greater gains in sensitivity ($73\%$ and $55\%$ in the two linear ranges compared to $26\%$  at the third cycle in CV). Anodized bulk copper electrodes in the literature (refer to Table 3 in Giziński et al. \cite{Stepniowski2020}) perform similarly with respect to glucose sensing. However, such a comparison would not be fair for several reasons. First, the resistivity of thin Cu films surpasses that of bulk Cu \cite{Cui_Hutt_Conway_2012}. Secondly, although highly scalable, the RF magnetron sputtering in the absence of ultra-high vacuum can lead to oxygen impurities in the film that increase its overall resistivity \cite{Lee_Kim_Lee_Kim_Kim_Park_Bae_Cho_Kim_Oh_etal._2014}. Lastly, a Pt sub-layer was introduced to ensure a dependable electrical connection. Thus, the electron transfer is further hampered by the inherently worse condution of Pt (compared to Cu) and the existence of a the Pt-Cu interface. Nonetheless, moving from bulk to film, together with some associated drawbacks, is a necessary step to achieve fully integrated and compact devices.
\subsubsection{Interference resistance}
\begin{figure} [h]
\centering
\includegraphics[width=0.5\linewidth]{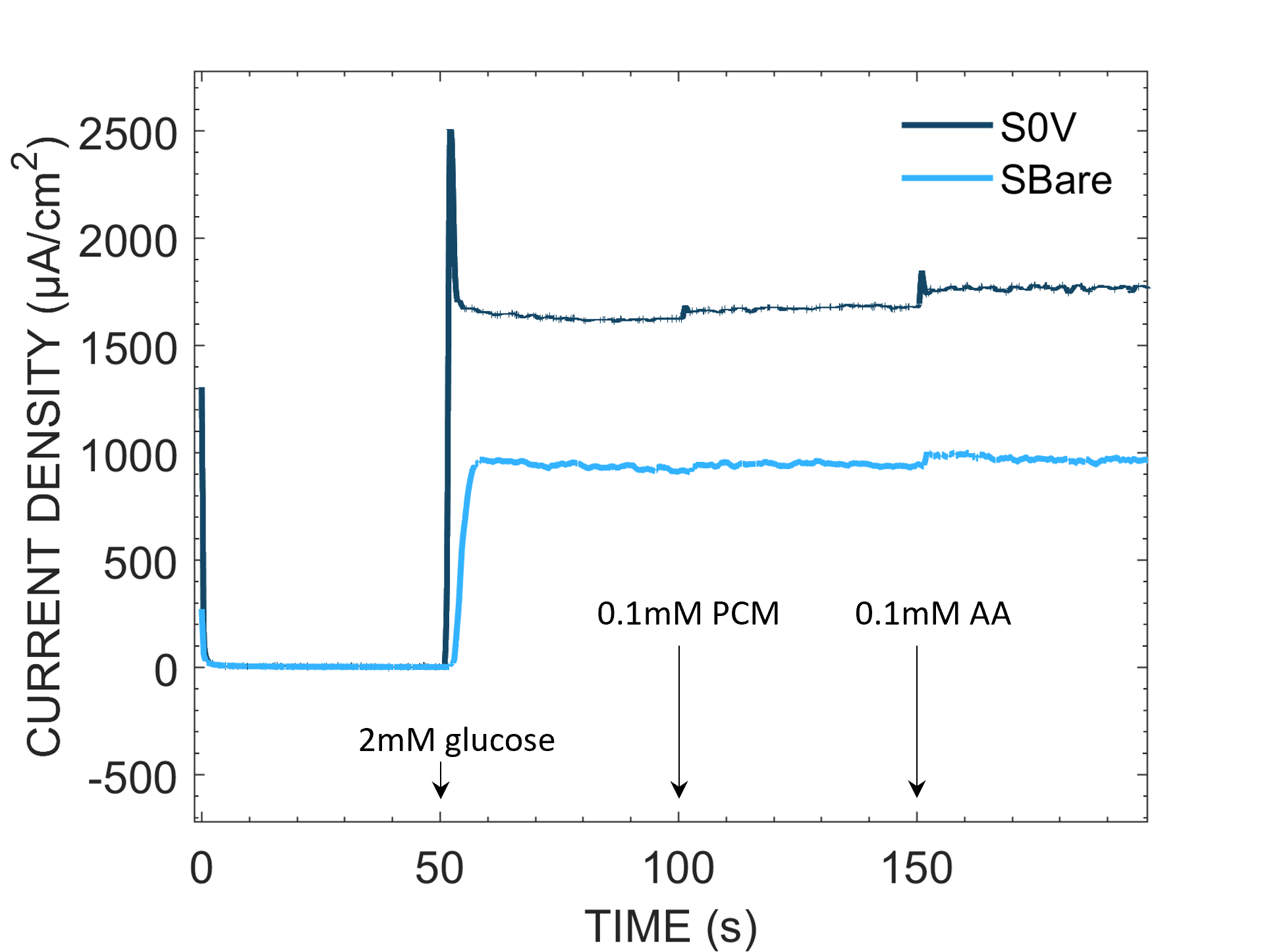}
\caption{Interference study at a potential of 0.5 V, with additions of 2 mM glucose, 0.1 mM ascorbic acid, 0.1 mM paracetamol. SBare is compared with S0V.}
\label{fig:interference}
\end{figure}

The presence of interferents in the blood serum, such as the easily oxidizable endogenous compounds like ascorbic acid (AA) and exogenous ones like paracetamol (PCM), may complicate the accurate estimation of glucose concentration\cite{Naikoo2022}. Thus, we evaluated the selectivity of our microfabricated Cu electrodes to glucose by chronoamperometry at an applied potential of 0.5 V. We added 2 mM of glucose to a 0.1 M KOH solution and then observed the change in current density at the addition of 0.1 mM of ascorbic acid first and of 0.1 mM acetaminophen thereafter (\textbf{Figure \ref{fig:interference}}). The relative concentrations were chosen to mimick blood serum values\cite{Wei_Qiao_Zhao_Liang_Li_Luo_Lu_Shi_Lu_Sun_2020}.
\noindent The addition of 0.1 mM of PCM caused a current density increase of $3\%$ and $4\%$ for S0V and SBare, respectively. Whereas, the addition of 0.1 mM of AA led to a $5\%$ and a $6\%$ increase for S0V and SBare, respectively. In conclusion, both electrodes showed satisfactory selectivity towards AA and PCM, with the anodized electrode having between 30$\%$ to 40$\%$ lower relative current change compared to the bare.

\section{Conclusions}
This study systematically delved into the relationship between anodization parameters and glucose oxidation on microfabricated copper electrodes, shedding light on their optimization for enhanced catalytic performance. This exploration culminated in the development of a three-dimensional plot, with glucose activity depicted as a function of the anodization potential and KOH concentration. The investigation is underpinned by a robust theoretical framework that complements the empirical findings, encompassing both the anodization process itself and the subsequent glucose electrooxidation. By precisely controlling the anodization conditions, specific copper oxide species can be selectively favored, namely $\mathrm{Cu_2O}$, $\mathrm{Cu(OH)_2}$, and CuO, each exhibiting varying catalytic activities. We demonstrated that the formation of copper hydroxide nanowires does not improve the catalytic activity towards glucose, even if accompanied by a substantial surface area increase. In contrast, tuning the anodization treatment parameters towards CuO formation was shown to enhance the catalytic activity of the thin film. Notably, the application of a polarization at 1 M KOH and 0 V (vs Ag\textbar AgCl) led to the formation of a highly active surface CuO layer, resulting in a significant improvement in sensitivity compared to bare copper electrodes. Specifically, yielding with chronoamperometry a 55 \% sensitivity gain in the 0.1 mM - 0.5 mM range, and an impressive 73 \% gain in the 0.75 mM - 2mM range. Our results offer a compelling evidence for the latest model by the group of Kubota, whereby the glucose oxidation activity of Cu electrodes is driven by the semiconductive properties of CuO and not by the Cu(II)/Cu(III) couple.
Furthermore, the produced electrodes exhibit an very low LOD – a mere 0.004 mM – which stands among the lowest reported in existing literature.  This minimal LOD can be attributed to the employed microfabrication technique that allows a cleanroom-compatible platform, while ensuring the scalability and reproducibility of manufacturing. By leveraging these insights, researchers and engineers can enhance the efficiency and performance of (bio)sensors and energy conversion devices by integrating a simple anodization step in the fabrication process. High surface area copper-based electrodes, such as thin films on nanostructured supports, will also greatly benefit from these findings. Further work will focus on exploring a greater range of potentials ($>$ 0 V) and KOH molarity ($>$ 1 M) in the direction of maximal efficiency for glucose oxidation, or by varying the anodization temperature to achieve pure CuO nanowires.
\clearpage

\section{TOC Graphic}
\begin{figure}
    \centering
    \includegraphics[width=\linewidth]{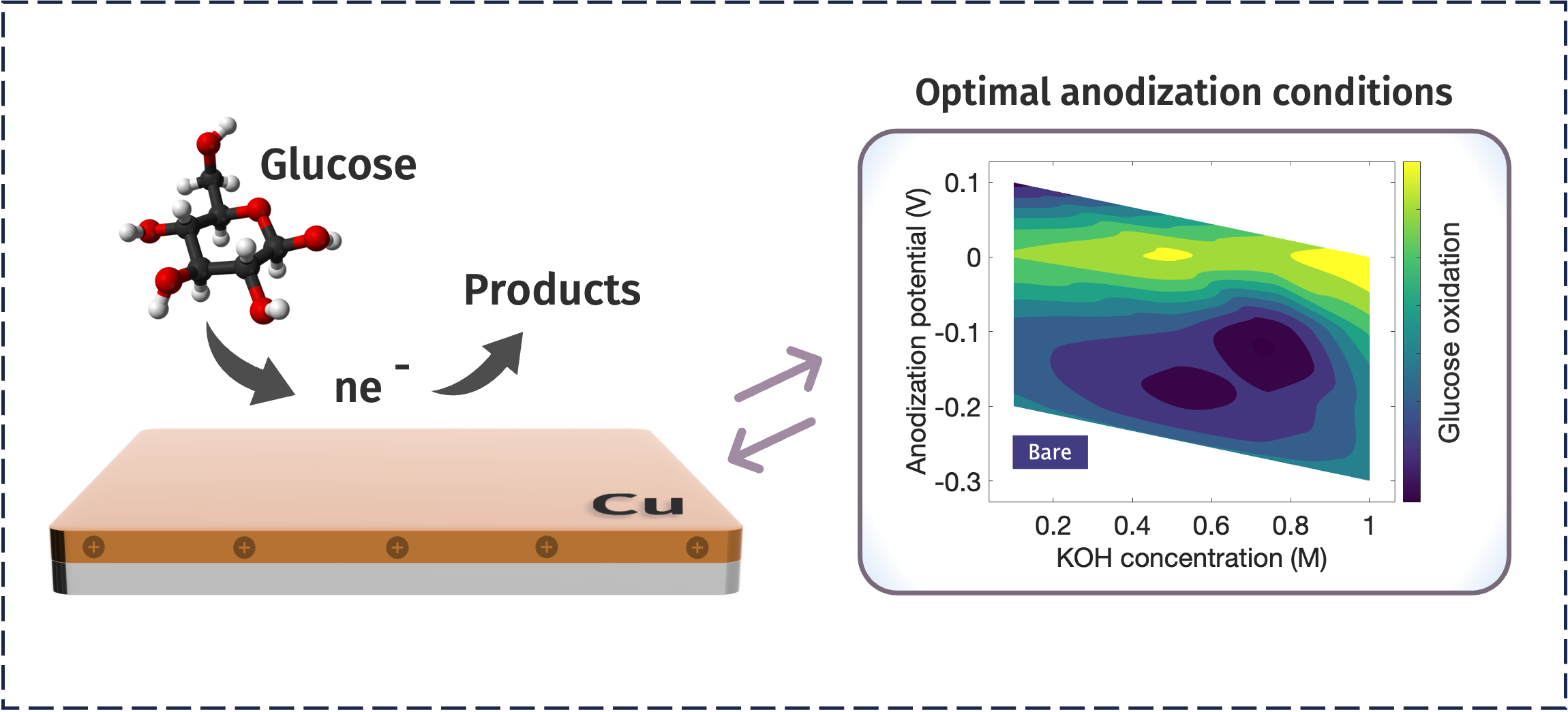}
\end{figure}
The relationship between anodization parameters and catalytic activity is investigated on microfabricated copper electrodes. A clear path to optimizing glucose oxidation through controlled anodization is presented, noting the formation of a highly active surface CuO layer in 1M KOH at 0 V (vs AgAgCl). These insights offer an intriguing solution to enhancing copper-based integrated devices for biosensing and energy conversion.

\clearpage
\bibliography{bibliography}

\clearpage
\begin{suppinfo}
\renewcommand{\figurename}{S}
\renewcommand{\tablename}{S Table}
\setcounter{figure}{0}
\setcounter{table}{0}

\begin{figure} [h!]
\centering
\includegraphics[width=\linewidth]{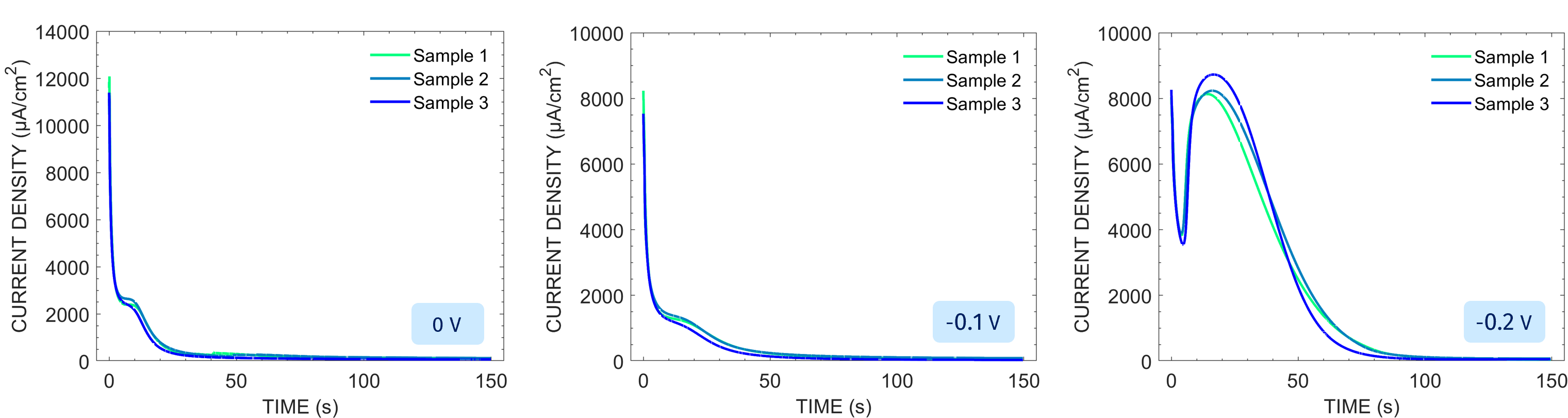}
\caption{Reproducibility study of anodizations performed three times on different electrodes at 0V, -0.1V and -0.2V in 1 M KOH.}
\label{fig:anodization-reprod}
\end{figure}

\begin{table}[]
\caption{RF sputtering parameters for Cu on Pt films }
\label{tab:fab}
\begin{tabular}{|lllll|}
\hline
\multicolumn{5}{|l|}{Base Pressure: 2.5E-05 mBar}                                                                                                               \\ \hline 
\multicolumn{1}{|l|}{Target} & \multicolumn{1}{l|}{Time (min)} & \multicolumn{1}{l|}{Power (W)} & \multicolumn{1}{l|}{Working pressure (mBar)} & Thickness (nm) \\ \hline \hline 
\multicolumn{1}{|l|}{Ti}     & \multicolumn{1}{l|}{0.5}        & \multicolumn{1}{l|}{200}       & \multicolumn{1}{l|}{1.00E-02}                & 10             \\ \hline
\multicolumn{1}{|l|}{Pt}     & \multicolumn{1}{l|}{7}          & \multicolumn{1}{l|}{100}       & \multicolumn{1}{l|}{1 E-02}                  & 300            \\ \hline
\multicolumn{1}{|l|}{Cu}     & \multicolumn{1}{l|}{5}          & \multicolumn{1}{l|}{200}       & \multicolumn{1}{l|}{1 E-02}                  & 300            \\ \hline
\end{tabular}
\end{table}

\begin{table}[]
\caption{XPS fitting parameters for SBare}
\label{tab:XPS-SBare}
\begin{tabular}{|l|l|l|l|l|}
\hline
\textbf{Component}     & \textbf{Position(eV)} & \textbf{FWHM(eV)} & \textbf{Lineshape} & \textbf{\%Area} \\ \hline \hline
Cu(OH)\textsubscript{2} & 934.57   & 2.95  & GL(30)    & 0.00   \\ \hline
Cu(OH)\textsubscript{2} & 939.20   & 2.90  & GL(30)    & 0.00   \\ \hline
Cu(OH)\textsubscript{2} & 942.10   & 2.80  & GL(30)    & 0.00   \\ \hline
Cu(OH)\textsubscript{2} & 944.02   & 1.66  & GL(30)    & 0.00   \\ \hline
Cu 2+                  & 933.21   & 2.11  & GL(30)    & 27.85  \\ \hline
Cu 2+                  & 934.58   & 3.01  & GL(30)    & 30.64  \\ \hline
Cu 2+ \tiny{(shakeup)} & 940.62   & 1.06  & GL(30)    & 2.23   \\ \hline
Cu 2+ \tiny{(shakeup)} & 941.75   & 3.64  & GL(30)    & 23.40  \\ \hline
Cu 2+ \tiny{(shakeup)} & 943.80   & 1.27  & GL(30)    & 5.29   \\ \hline
Cu 1+                  & 932.27   & 1.10  & GL(80)    & 10.55  \\ \hline
Cu 0+                  & 932.70   & 0.90  & GL(90)    & 0.05   \\ \hline
\end{tabular}
\end{table}

\begin{table}[]
\caption{XPS fitting parameters for S0V}
\label{tab:XPS-S0V}
\begin{tabular}{|l|l|l|l|l|}
\hline
\textbf{Component}     & \textbf{Position(eV)} & \textbf{FWHM(eV)} & \textbf{Lineshape} & \textbf{\%Area} \\ \hline \hline
Cu(OH)\textsubscript{2} & 934.77   & 2.95  & GL(30)    & 3.40   \\ \hline
Cu(OH)\textsubscript{2} & 939.40   & 2.70  & GL(30)    & 0.34   \\ \hline
Cu(OH)\textsubscript{2} & 942.30   & 3.00  & GL(30)    & 1.19   \\ \hline
Cu(OH)\textsubscript{2} & 944.22   & 1.66  & GL(30)    & 0.31   \\ \hline
Cu 2+                  & 933.19   & 2.20  & GL(30)    & 29.09  \\ \hline
Cu 2+                  & 934.56   & 3.15  & GL(30)    & 32.00  \\ \hline
Cu 2+ \tiny{(shakeup)} & 940.60   & 1.09  & GL(30)    & 2.33   \\ \hline
Cu 2+ \tiny{(shakeup)} & 941.73   & 3.65  & GL(30)    & 24.44  \\ \hline
Cu 2+ \tiny{(shakeup)} & 943.78   & 1.26  & GL(30)    & 5.53   \\ \hline
Cu 1+                  & 932.37   & 0.90  & GL(80)    & 1.13   \\ \hline
Cu 0+                  & 932.53   & 0.70  & GL(90)    & 0.26   \\ \hline
\end{tabular}
\end{table}

\begin{table}[]
\caption{XPS fitting parameters for S-0.1V}
\label{tab:XPS-S-0.1V}
\begin{tabular}{|l|l|l|l|l|}
\hline
\textbf{Component}     & \textbf{Position(eV)} & \textbf{FWHM(eV)} & \textbf{Lineshape} & \textbf{\%Area} \\ \hline \hline
Cu(OH)\textsubscript{2} & 934.72   & 2.95  & GL(30)    & 37.70  \\ \hline
Cu(OH)\textsubscript{2} & 939.35   & 2.70  & GL(30)    & 3.77   \\ \hline
Cu(OH)\textsubscript{2} & 942.25   & 3.00  & GL(30)    & 13.20  \\ \hline
Cu(OH)\textsubscript{2} & 944.17   & 1.66  & GL(30)    & 3.39   \\ \hline
Cu 2+                  & 933.09   & 2.20  & GL(30)    & 12.86  \\ \hline
Cu 2+                  & 934.46   & 3.15  & GL(30)    & 14.15  \\ \hline
Cu 2+ \tiny{(shakeup)} & 940.50   & 1.13  & GL(30)    & 1.03   \\ \hline
Cu 2+ \tiny{(shakeup)} & 941.63   & 3.65  & GL(30)    & 10.80  \\ \hline
Cu 2+ \tiny{(shakeup)} & 943.68   & 1.27  & GL(30)    & 2.44   \\ \hline
Cu 1+                  & 932.08   & 1.10  & GL(80)    & 0.15   \\ \hline
Cu 0+                  & 932.73   & 0.90  & GL(90)    & 0.51   \\ \hline
\end{tabular}
\end{table}

\begin{table}[]
\caption{XPS fitting parameters for S-0.2V}
\label{tab:XPS-S-0.2V}
\begin{tabular}{|l|l|l|l|l|}
\hline
\textbf{Component}     & \textbf{Position(eV)} & \textbf{FWHM(eV)} & \textbf{Lineshape} & \textbf{\%Area} \\ \hline \hline
Cu(OH)\textsubscript{2} & 934.66   & 2.95  & GL(30)    & 0.54   \\ \hline
Cu(OH)\textsubscript{2} & 939.29   & 2.90  & GL(30)    & 0.05   \\ \hline
Cu(OH)\textsubscript{2} & 942.19   & 3.00  & GL(30)    & 0.19   \\ \hline
Cu(OH)\textsubscript{2} & 944.11   & 1.66  & GL(30)    & 0.05   \\ \hline
Cu 2+                  & 933.21   & 2.18  & GL(30)    & 26.31  \\ \hline
Cu 2+                  & 934.58   & 2.95  & GL(30)    & 28.94  \\ \hline
Cu 2+ \tiny{(shakeup)} & 940.62   & 1.05  & GL(30)    & 2.10   \\ \hline
Cu 2+ \tiny{(shakeup)} & 941.75   & 3.65  & GL(30)    & 22.10  \\ \hline
Cu 2+ \tiny{(shakeup)} & 943.80   & 1.27  & GL(30)    & 5.00   \\ \hline
Cu 1+                  & 932.25   & 1.10  & GL(80)    & 10.29  \\ \hline
Cu 0+                  & 932.53   & 0.80  & GL(90)    & 4.42   \\ \hline
\end{tabular}
\end{table}

\begin{figure} [h!]
\centering
\includegraphics[width=\linewidth]{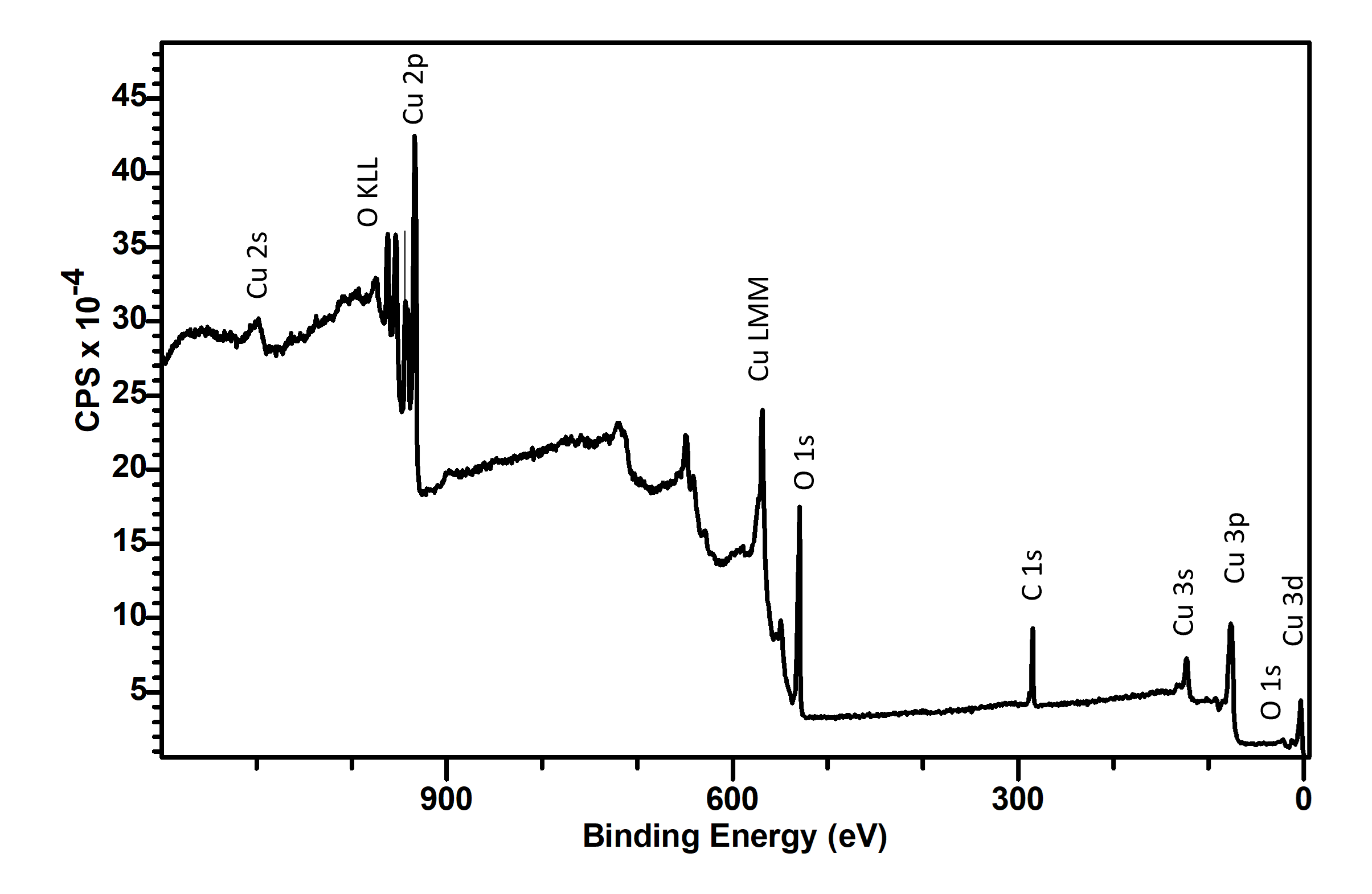}
\caption{XPS survey spectrum of the fabricated bare copper electrode}
\label{fig:surveyXPS}
\end{figure}

\begin{figure} [h!]
\centering
\includegraphics[width=\linewidth]{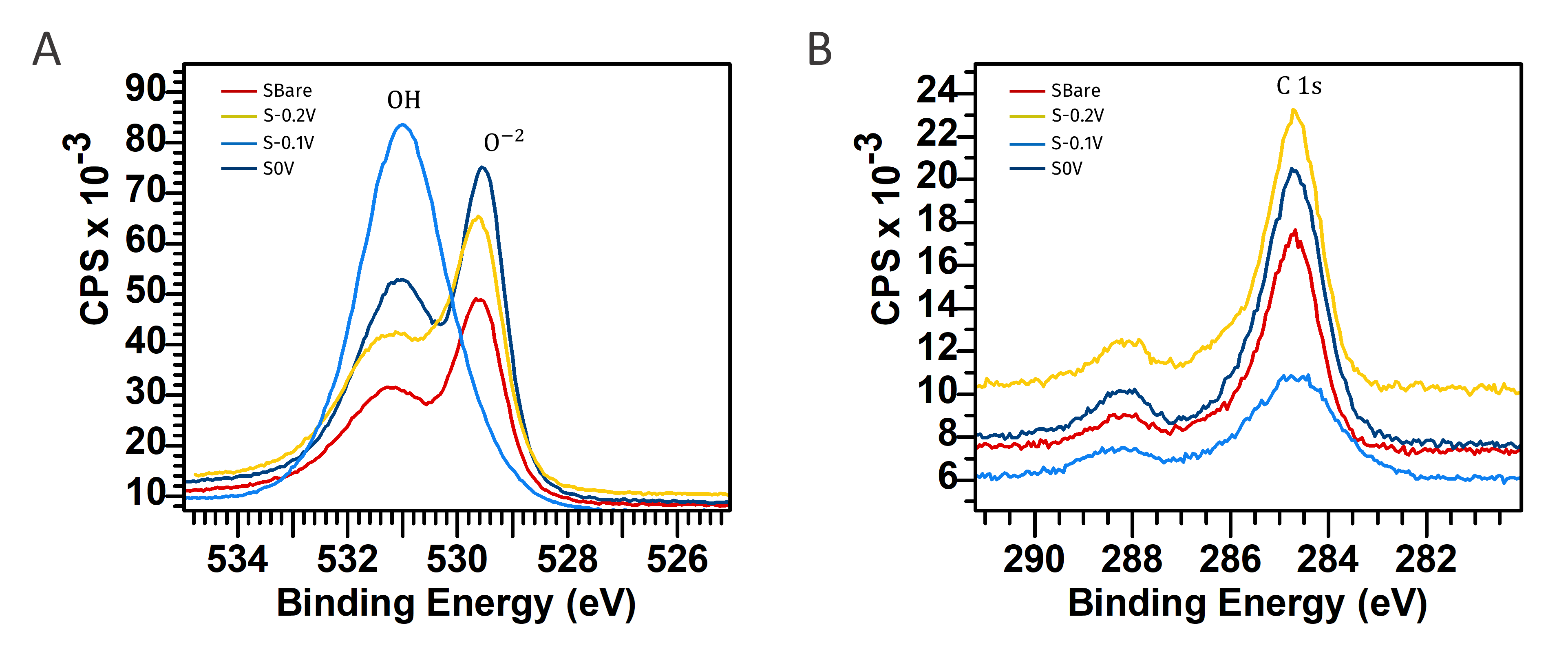}
\caption{(A) O 1s XPS peaks for sample SBare, S0V, S-0.1V, S-0.2V.(B) C 1s XPS peaks for sample SBare, S0V, S-0.1V, S-0.2V }
\label{fig:O1sC1s}
\end{figure}

\begin{figure} [h!]
\centering
\includegraphics[width=\linewidth]{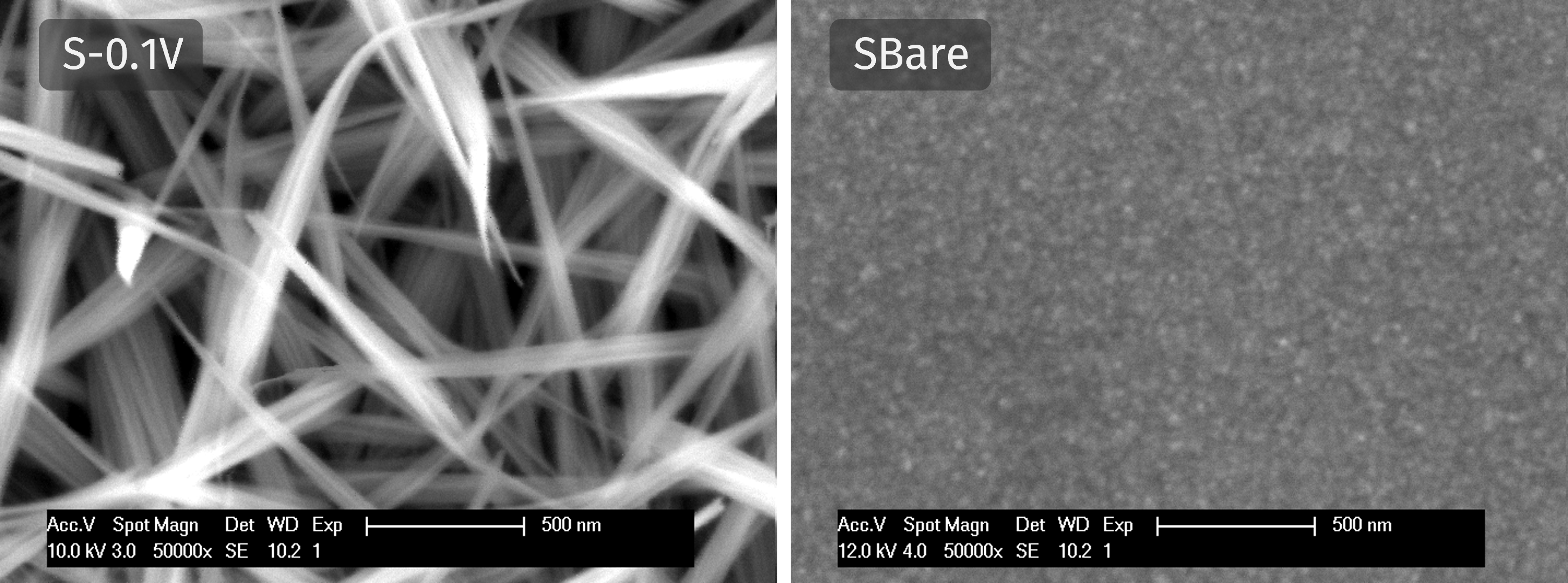}
\caption{SEM images of S0V (left) and SBare (right), with a magnification of 50kX}
\label{fig:SEMzoom}
\end{figure}

\begin{figure} [h!]
\centering
\includegraphics[width=0.5\linewidth]{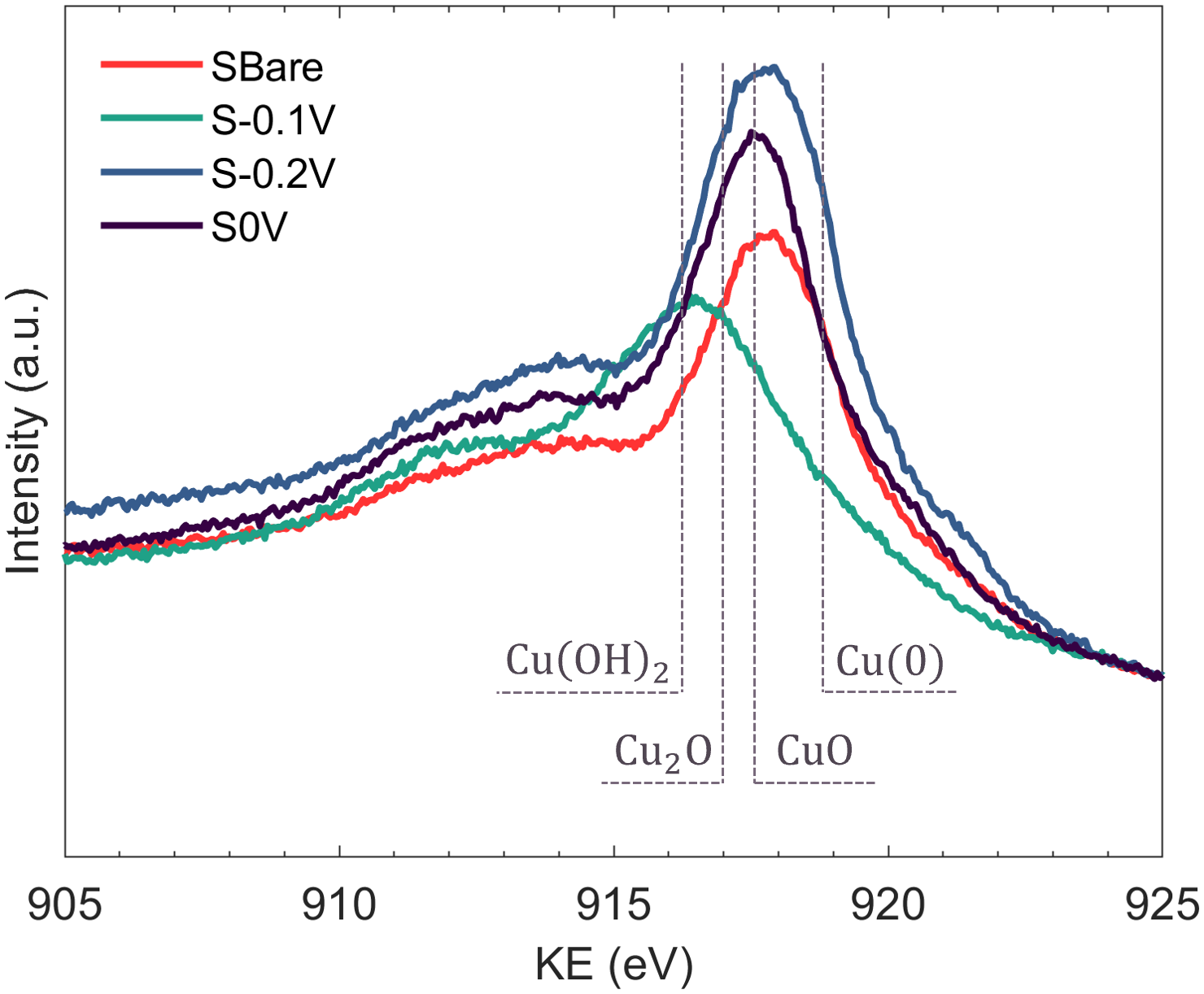}
\caption{Cu $\mathrm{L_{3}M_{45}M_{45}}$ Auger spectra of the samples anodized at various potentials}
\label{fig:CuLMM}
\end{figure}

\begin{figure} [h!]
\centering
\includegraphics[width=\linewidth]{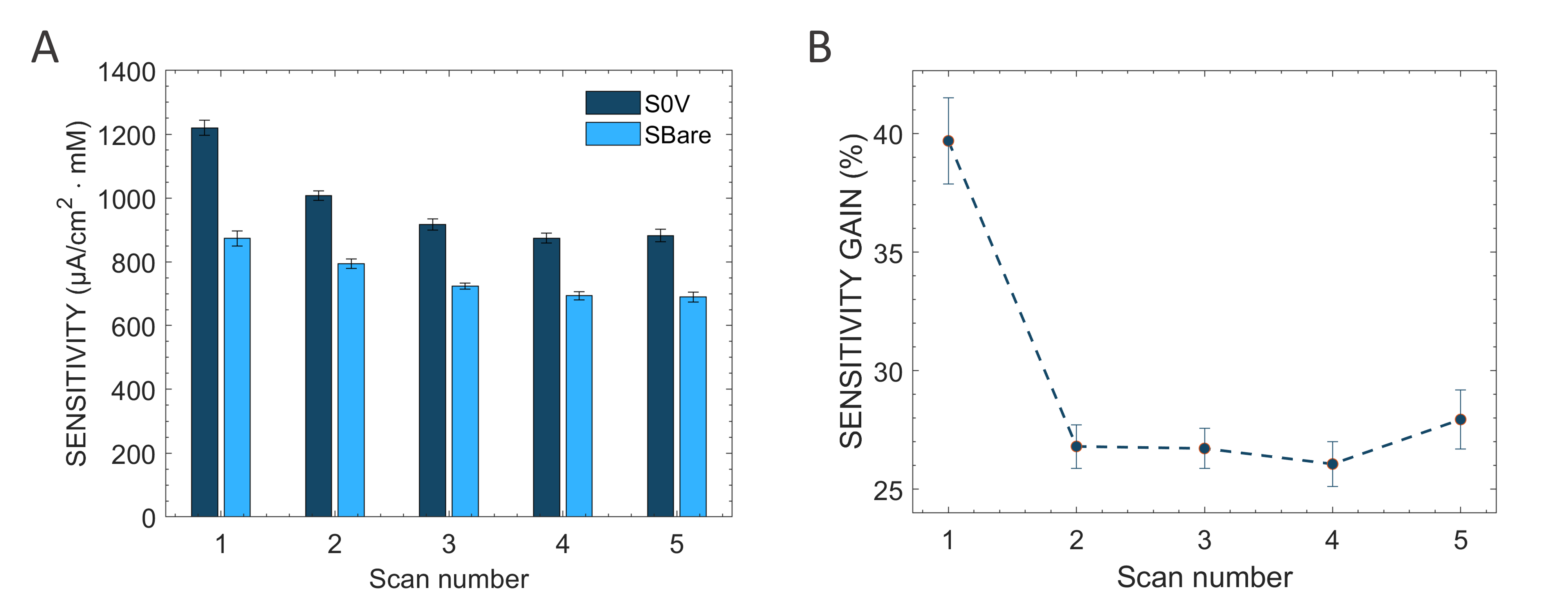}
\caption{(A) Calculated sensitivity based on cyclic voltammetry of S0V and SBare as a function of the scan number. (B) Sensitivity gain of S0V compared to Sbare as a function of the scan number}
\label{fig:sensitivity-scan}
\end{figure}

\begin{table}[]
\caption{Experimental data used to built a contour plot}
\label{tab:countourdata}
\begin{tabular}{|c|c|c|}
\hline
\multicolumn{1}{|l|}{\textbf{KOH (M)}} & \multicolumn{1}{l|}{\textbf{Potential vs AgAgCl (V)}} & \multicolumn{1}{l|}{\textbf{Peak current CV ($\mathrm{\mu}$A/cm\textasciicircum{}2)}} \\ \hline
0.1                                    & 0.1                                                   & 4816.71                                                                               \\ \hline
0.1                                    & -0.1                                                  & 5804.63                                                                               \\ \hline
0.1                                    & -0.2                                                  & 6260.65                                                                               \\ \hline
0.1                                    & 0                                                     & 7501.61                                                                               \\ \hline
0.5                                    & -0.1                                                  & 5495.08                                                                               \\ \hline
0.5                                    & -0.2                                                  & 5011.04                                                                               \\ \hline
0.5                                    & 0                                                     & 8034.39                                                                               \\ \hline
0.75                                   & -0.1                                                  & 4624.61                                                                               \\ \hline
0.75                                   & -0.2                                                  & 5153.31                                                                               \\ \hline
0.75                                   & 0                                                     & 7895.24                                                                               \\ \hline
1                                      & -0.1                                                  & 7095.90                                                                               \\ \hline
1                                      & -0.2                                                  & 6232.71                                                                               \\ \hline
1                                      & -0.3                                                  & 6494.98                                                                               \\ \hline
1                                      & 0                                                     & 8270.49                                                                               \\ \hline
\end{tabular}
\end{table}

\begin{figure} [h!]
\centering
\includegraphics[width=\linewidth]{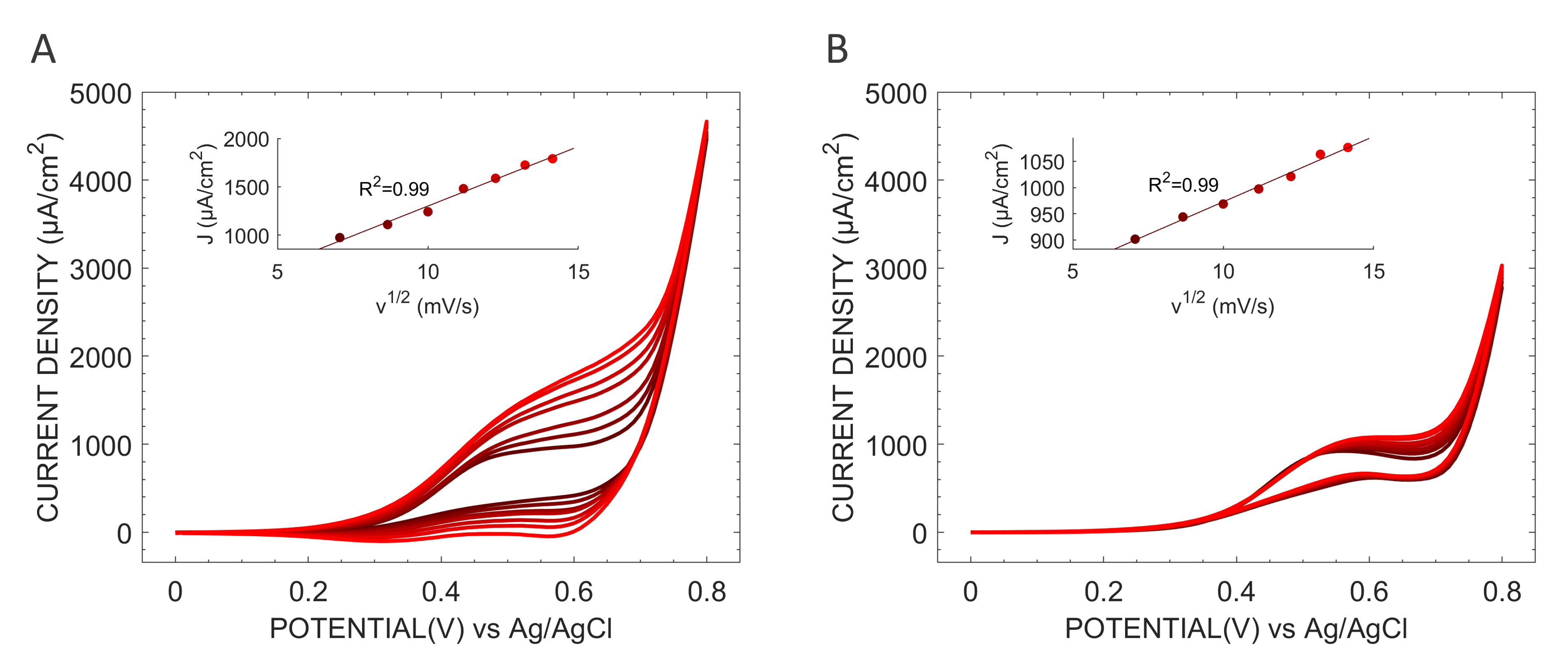}
\caption{Scan rate study for a 0V anodized electrode (A) and the bare copper electrode (B). The third cycle is taken and the current densities at 0.6 V are considered. Glucose concentration: 2 mM}
\label{fig:scanrate}
\end{figure}

\end{suppinfo}

\end{document}